\documentclass[journal=jacsat,manuscript=article]{achemso}

\usepackage{chemformula}
\usepackage[T1]{fontenc}

\author{Daniel Graf}
\email{dg641@cam.ac.uk}
\author{Alex J. W. Thom}
\affiliation[University of Cambridge]
{Yusuf Hamied Department of Chemistry, University of Cambridge, Cambridge}

\title[DC(HF)-DFT]
  {A simple and efficient route towards improved energetics within the 
  framework of 
  density-corrected density functional theory}

\keywords{density-corrected density functional theory, dc-DFT, HF-DFT}

\begin{document}

\begin{tocentry}
    \includegraphics[width=8.3cm]{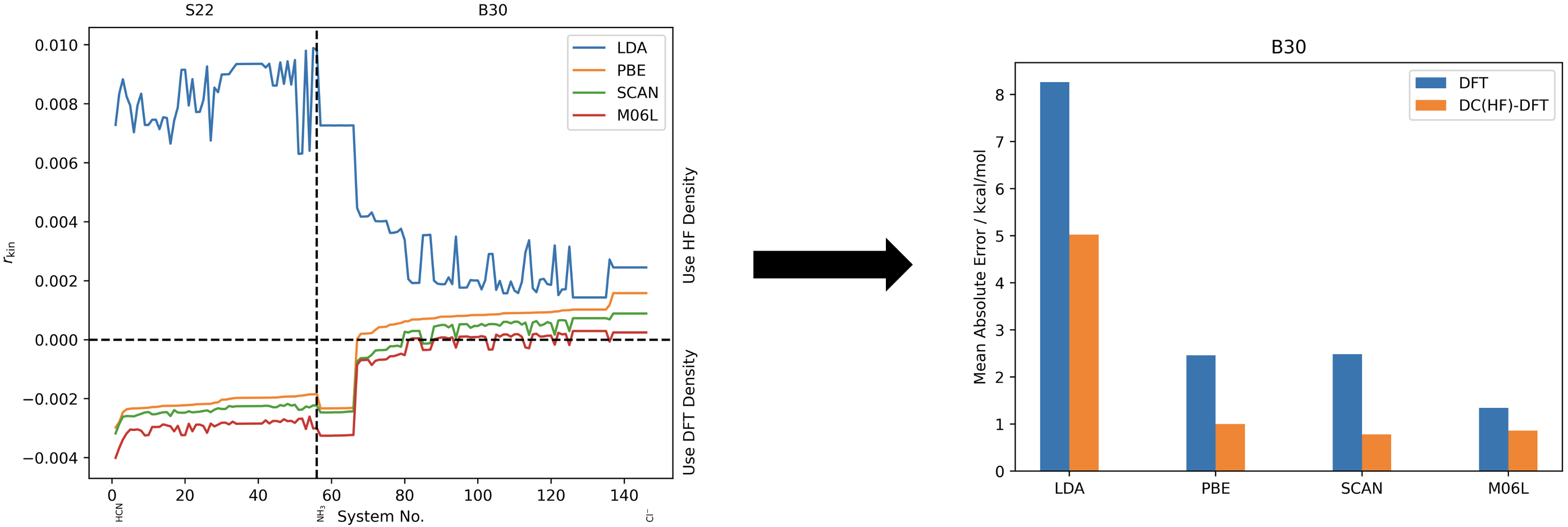}
\end{tocentry}

\begin{abstract}
    The crucial step in density-corrected Hartree--Fock density functional
    theory (DC(HF)-DFT) is to decide whether the density produced by the
    density functional for a specific calculation is erroneous and hence should
    be replaced by, in this case, the HF density. 
    We introduce an indicator, based on the difference in non-interacting
    kinetic energies between DFT and HF calculations, to determine when the HF
    density is the better option.
    Our kinetic energy indicator directly compares the self-consistent density 
    of the analysed functional with the HF density, is size-intensive, 
    reliable, and most importantly highly efficient.  

    Moreover, we present a procedure that makes best use of the computed
    quantities necessary for DC(HF)-DFT by additionally evaluating a related
    hybrid functional and, in that way, not only ``corrects'' the density but
    also the functional itself; we call that procedure corrected Hartree--Fock
    density functional theory (C(HF)-DFT).
\end{abstract}

\section{Introduction}
Density functional theory (DFT) is a widely used approach in computational 
physics and chemistry, owing to the fact that it allows for the relatively 
simple approximation of many-body effects, providing useful accuracy at low 
computational cost. Despite the existence of hundreds of density functionals, 
most DFT calculations use only a few standard functionals, often in the form 
of (meta) general gradient approximations ((m)GGAs).\cite{vuckovic2019}
While (m)GGAs are true Kohn--Sham\cite{kohn1965} (KS) density functionals, 
consisting of a 
local multiplicative KS potential, local and semi-local density functionals 
tend to over-delocalise charge. This over-delocalisation is associated with 
several well-known problems in density functional theory, 
including delocalisation error,\cite{sanchez2008,
cohen2007, cohen2008, li2017, li2015, johnson2013, vazquez2015, leblanc2018} 
one-electron self-interaction error,\cite{perdew1981} many-electron 
self-interaction error,\cite{sanchez2006, vydrov2007, ruzsinszky2007} missing 
derivative discontinuities in the energy as particle numbers pass through 
integer values --- density functionals are too smooth ---\cite{perdew1982, 
sanchez2009, yang2012} and fractional charge and spin errors;\cite{zhang1998, 
ruzsinszky2006, cohen2014} and is the reason for e.g. unbound anions, 
incorrect molecular dissociation curves, and underestimated reaction 
barriers.\cite{cohen2008, cohen2012}

To address the problem of over-delocalisation, various approaches have been 
developed, such as self-interaction corrections,\cite{perdew1981} 
the admixture of exact Hartree--Fock\cite{hartree1928, slater1930, fock1930}
(HF) exchange,
the localised orbital scaling correction (LOSC),\cite{li2017} and 
range-separation methods\cite{leininger1997}. 
Moreover, in cases where standard density functionals fail, using the HF 
density instead of 
the self-consistent density, known as HF-DFT, has been shown to 
improve results significantly.\cite{gordon1972, colle1975, scuseria1992, 
janesko2008, cioslowski1993, oliphant1994, verma2012}
For a comprehensive benchmark of HF-DFT the interested reader is referred to the
work of
Martin and co-workers\cite{santra2021}.

The good performance of HF-DFT and its appealing theoretical and practical
simplicity has led Burke and co-workers to the development of density-corrected 
(HF) density functional theory (DC(HF)-DFT).\cite{nam2021, nam2020, sim2018,
sim2022, vuckovic2019, song2021, kim2013, martin2021, kim2019, kim2014,
wasserman2017}
Broadly speaking, this method involves two key
steps: assessing whether the density generated by the density functional 
requires correction or replacement, and then, \textit{if necessary}, 
substituting the HF density and 
evaluating the functional on that density (performing a HF-DFT calculation). 
This strategy sets DC(HF)-DFT apart from
pure HF-DFT, as it ensures --- at least in theory --- that the HF density is 
used only when it improves
the accuracy of the results. While DC(HF)-DFT has already demonstrated great
potential,\cite{kim2011, kim2015, nam2020, kim2014, kim2019, song2018,
lee2010b, lee2010, nam2021, lambros2021, dasgupta2021} in this work we show 
that further 
enhancements are possible.

\section{Theoretical considerations}
\subsection{Why density-corrections might be necessary and useful}
The exchange-correlation functional is the only part of (KS-)DFT that is not 
known 
exactly and hence needs to be approximated. This approximation is then used 
twice in common DFT calculations, once when determining the density and 
again when determining the energy of the system; of course, neither is 
exact. 
Despite the name, the accuracy of a certain density functional in terms of 
energetics does not necessarily 
guarantee the accuracy of the KS potential or the density itself.
In fact, most density functional approximations (DFAs) produce poor 
KS potentials\cite{umrigar1994, cruz1998} which can be seen e.g. in the 
poor orbital energies\cite{kuemmel2008} these functionals yield. 
Nevertheless, in most cases, the density is still very accurate\cite{kim2013} 
because the overall shape of the approximate potential is reasonable, 
although it is shifted with respect to the exact one, which does not affect 
the orbitals or the density.\cite{kim2014, wasserman2017} 

However, there are large classes of calculations where the density is poor, 
leading to significant errors in the calculated energies.\cite{sim2018,
nam2020, vuckovic2019, song2021} 
Burke and co-workers developed a framework to distinguish such 
\textit{density-driven errors} 
from the errors of the functional itself,\cite{vuckovic2019} 
the \textit{functional errors}, by 
separating the total error according to
\begin{equation}
    \Delta \tilde{E} = \underbrace{\tilde{E} [\tilde{n}] - 
    E [\tilde{n}]}_{\Delta \tilde{E}_{\text{xc}} [\tilde{n}]}
    + \underbrace{E [\tilde{n}] - E [n]}_{D^{\text{ideal}} [\Delta
    \tilde{n}]} 
    \label{eq:e_sep_ideal}
\end{equation}
where exact quantities are denoted without a tilde while approximate quantities 
are denoted with a tilde symbol; e.g. $E[\tilde{n}]$
denotes the exact functional evaluated on an approximate density.
Since it is impractical to evaluate the exact functional on an approximate 
density, the following separation was proposed:
\begin{equation}
    \Delta \tilde{E} = \underbrace{\tilde{E} [\tilde{n}] - 
    \tilde{E} [n]}_{D^{\text{approx}} [\Delta \tilde{n}]}
    + \underbrace{\tilde{E} [n] - E [n]}_{\Delta E_{\text{F}}}
\end{equation}
where the density-driven error ($D^{\text{approx}}$) is now obtained using an
approximate functional $\tilde{E}$.
If the density-driven error exceeds the 
functional error ($\Delta E_{\text{F}}$), the calculation is 
considered \textit{abnormal}, which means that the functional itself is (or 
can be) accurate while the produced density, due to a wrong potential, is 
poor.\cite{burke1998} 
For a more detailed discussion of how this is possible and the underlying 
theory in general, the reader is referred to Ref.~\citenum{vuckovic2019}.

Although highly accurate densities can be computed using coupled cluster or 
configuration interaction approaches, 
there are differences in the correlation energy in wave-function theory and 
density functional theory.
Moreover, these high-level wave-function methods produce interacting 
kinetic energies, while the KS framework requires non-interacting ones. 
To obtain a corresponding local KS potential and its associated KS orbitals and 
orbital energies, the density must be inverted, which is computationally 
expensive and numerically challenging.\cite{nam2020, kaplan2023}
However, it has been shown that for abnormal calculations  --- contrary to 
normal calculations, where the functional error dominates --- the use of the 
HF density is, in terms of improving the energetics, often not very different 
from the use of the exact (or highly accurate) density.\cite{sim2022} 
We want to stress that this does not necessarily mean that the HF density is 
overall better because there is no well-defined meaning of a better
density,\cite{mezei2017, medvedev2017, brorsen2017, sim2018,
mostafanejad2019, medvedev2017b, hammes2017, kepp2017, gould2017, mayer2017} 
as pointed out by Burke and co-workers several times. 
It simply means that the density functional evaluated on the HF density shows
a smaller density-driven error in these cases.\cite{sim2022}

As previously mentioned, the use of the HF density can be very beneficial; 
nevertheless, we may not always want to use the HF density. 
First of all, self-consistency makes the evaluation of properties depending on 
the derivative of the energy much easier to calculate since a lot of terms 
vanish. However, we note that a scheme of 
calculating gradients for HF-DFT was put forward by Bartlett and 
co-workers.\cite{verma2012}
Moreover, for normal cases, the self-consistent density usually yields more
accurate energetics.\cite{kim2013}
And finally, the HF density should not be employed if it is spin-contaminated
since it should no longer be considered more accurate,
as pointed out by Burke and co-workers.\cite{song2022} 

\subsection{When to correct the density}
Recently, there has been a vigorous discussion about how to evaluate the 
accuracy of 
densities.\cite{mezei2017, medvedev2017, brorsen2017, sim2018,
mostafanejad2019, medvedev2017b, hammes2017, kepp2017, gould2017, mayer2017}
The problem with this is that the density is a function\cite{sim2018}, meaning 
that there are 
infinitely many numbers to compare and hence many ways to do so.
Burke and co-workers argued\cite{sim2018} that the energy is the most 
meaningful measure since it is the 
quantity that really matters and it is further able to detect even the tiniest
differences in the density \textit{when they matter}, leading to the 
development of density functional analysis.\cite{sim2022}
The present work deals with more pragmatic, but related, questions:
When is the HF density likely to improve the results obtained with a certain
density functional? And how can we decide that efficiently?

In order to detect abnormal calculations, Burke and co-workers put forward a 
simple heuristic called the \textit{density sensitivity} defined
as\cite{sim2022}
\begin{equation}
    \tilde{S} = \lvert \tilde{E} [ n^{\text{LDA}} ] -
    \tilde{E} [ n^{\text{HF}} ] \rvert
    \label{eq:dens_sens}
\end{equation}
where $n^{\text{LDA}}$ and $n^{\text{HF}}$ denote the LDA and the HF density,
respectively. Note that Eq.~\ref{eq:dens_sens} represents the density
sensitivity of \textit{one} calculation, but the density sensitivity is
usually evaluated for the whole reaction of interest. If the density
sensitivity of this reaction is above a certain threshold (2 kcal/mol 
is the usual choice\cite{sim2018}) 
the reaction is considered density sensitive and the HF density is employed
instead of the self-consistent density to evaluate the reaction energy.

Comparing Eq.~\ref{eq:dens_sens} with Eq.~\ref{eq:e_sep_ideal} it becomes 
apparent that this measure 
resembles the exact one if the curvature of the approximate functional 
is accurate,\cite{vuckovic2019} the LDA density is close to the self-consistent 
density of the functional under investigation (denoted by $\tilde{E}$), and 
the HF density is close to the exact one. These conditions are, of course,
rarely met, but this is not very problematic since we are only  
interested in answering the question whether the energy calculation is 
sensitive with
respect to the density in use or not. However, there are some
weaknesses of the proposed density sensitivity measure, especially in
combination with DC(HF)-DFT: 

First of all, the density sensitivity is independent of the density
generated by the functional being analysed, although it can differ 
significantly from the LDA one. 
Furthermore, when the density sensitivity exceeds a specified threshold, the
HF density is \textit{presumed} to be a better choice than the self-consistent 
density, or even an accurate approximation of the exact density.\cite{kim2019}  
This assumption, coupled with the utilization of the LDA density, 
rather than the functional's self-consistent density, 
introduces potential difficulties.

Moreover, the density sensitivity is size extensive, which
necessitates adjustment of the threshold according to the system 
size.\cite{martin2021, nam2021}
Additionally, when calculating small energy values such as torsional barrier
heights or non-covalent interactions, the threshold must be further 
adapted,\cite{song2022, nam2021} which can introduce an element of 
arbitrariness.

As mentioned above, the density sensitivity could, in principle, 
be applied to single 
calculations, but it is typically used for reaction energies. While it is, of
course, true that key chemical concepts are determined by energy differences
and that absolute energies are not even observables,\cite{kepp2018} 
this introduces a source of error cancellation.\cite{vuckovic2019}
There is a further source of error cancellation in the density
sensitivity measure: since the density sensitivity is measured using an
approximate exchange-correlation functional, errors in that functional can
cancel the ones in the density as functional errors and density-driven errors
have opposite signs.\cite{kim2014} That such an error cancellation can occur is
well known.\cite{crisostomo2022b, kaplan2023, janesko2008}

In that context, we also mention the work of Kepp, who proposed a 
recipe to assess the degree of normality which evaluates four distinct 
functionals on each other's self-consistent densities.\cite{kepp2018} 
The use of various functionals 
reduces the probability of error cancellation in measuring the 
abnormality of the reaction. However, the HF density was not included in this 
measure, preventing it from detecting a lot of abnormalities. 
Additionally, for a trial set of $N$ functionals,
$N^2$ calculations are necessary for each system, which is computationally
demanding.

This leads us to a final issue: the value of DFT 
lies in its computational efficiency, and this would be significantly reduced 
if additional HF 
calculations had to be performed every time. 
Since the majority of calculations are not density sensitive\cite{song2022},
this is a weakness needing to be addressed in 
order to facilitate more widespread use of DC(HF)-DFT.

In the subsequent discussion, we will try to address the aforementioned
weaknesses of the density sensitivity by proposing a novel simple and efficient
heuristic 
approach based on the non-interacting kinetic energy for detecting abnormal 
DFT calculations.

\section{The kinetic energy indicator}
\subsection{Theoretical rationalisation}
To begin with, let us summarise the key features that an indicator should 
possess in order to signal the superiority of the HF density for a given DFT
calculation, as these characteristics serve as 
the foundation for our kinetic energy indicator:

First, the indicator should compare, using a specified metric, 
the \textit{self-consistent density of the specific functional}  
with the HF density. 
Second, it should be \textit{size-intensive}; so, no adjustment of thresholds 
should be necessary.
Third, it should \textit{avoid error cancellation} as much as possible. 
Fourth, it should be \textit{efficient}.

Our proposed kinetic energy indicator is very simple and requires two
calculations: 
a converged DFT calculation using our preferred density functional and a
converged HF
calculation on the very same system; it then compares the two 
(non-interacting) kinetic energies. 
If the HF kinetic energy is larger than the one obtained from the DFT
calculation, the HF density is the better choice. But how did we
arrive at that conclusion? 

We first appeal to the textbook example of a
particle in a 1-dimensional box with potential $V (\mathbf{r} ) = 0$. Since we
set the potential to 0, the energy of the particle is given by\cite{pendas2022}
\begin{equation}
    E = T = \frac{\pi^2}{2 L^2}
\end{equation}
with $L$ denoting the length of the box. As is obvious, the total energy 
--- and hence the kinetic energy --- becomes smaller the larger the box gets. 
Transferring the conclusion from this extremely simplified example to the
problem of delocalisation, we would expect a similar behaviour: 
a decrease in the kinetic energy if the system delocalises. 

To illustrate the lowering of the kinetic energy with increasing
delocalisation, we calculated the energies of the H atom using different
functionals of the form
\begin{equation}
    E = T_{\text{s}} + E_{\text{en}} + E_{\text{J}} + 
    (1 - a) E^{\text{PBE}}_{\text{xc}} + a E^{\text{HF}}_{\text{x}}
    + E_{\text{nn}}
    \label{eq:pbex}
\end{equation}
where we vary the value of the mixing factor $a$ from 0 to 1. In
Eq.~\ref{eq:pbex}, $T_{\text{s}}$ denotes the non-interacting kinetic energy, 
$E_{\text{en}}$ denotes the energy stemming from the attraction of the
electrons to the nuclei, $E_{\text{J}}$ denotes the so-called Coulomb energy,
$E^{\text{PBE}}_{\text{xc}}$ denotes the PBE\cite{perdew1996, perdew1997}
exchange-correlation (xc) energy,
and $E^{\text{HF}}_{\text{x}}$ denotes the HF exchange energy. Note that
we scale the complete PBE xc-energy and so the PBE0\cite{adamo1999,
ernzerhof1999} functional is not within
the set of functionals, but we recover the standard PBE functional for $a = 0$ 
and the HF functional for $a = 1$.  
\begin{figure}
    \includegraphics{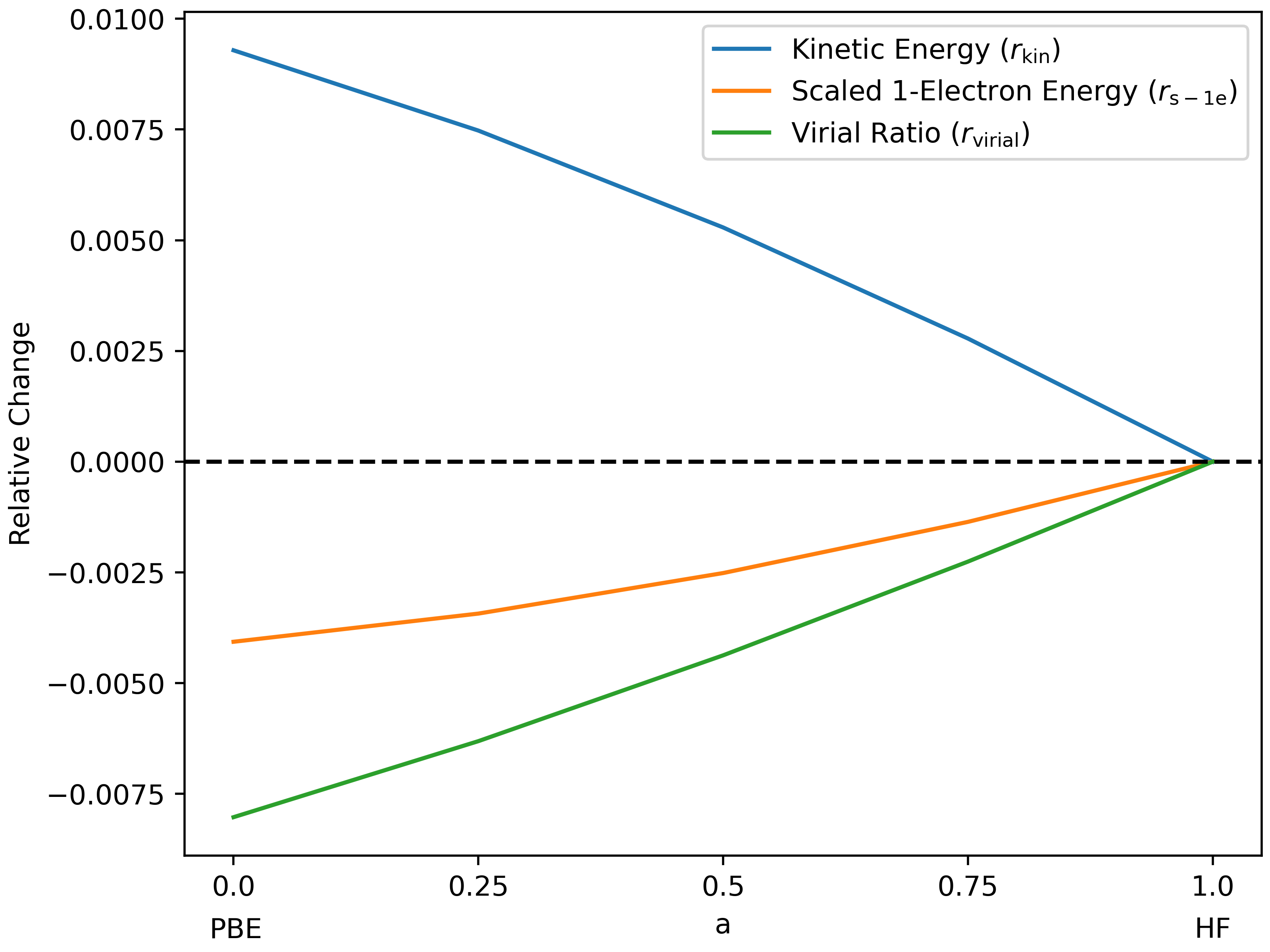}
    \caption{\label{fig:de_contribs_pbex_h_atom} Behaviour of the two
    indicators and the virial ratio difference when interpolating between 
    pure PBE and pure HF (exact) for the H atom. 
    Note that both the exchange and the correlation part of the
    PBE functional are
    scaled and hence the functional obtained for mixing factor $0.25$ does not
    correspond to the PBE0 functional.}
\end{figure}

The relative change of the kinetic energy is given by
\begin{equation}
    r_{\text{kin}} = \frac{T^{\text{HF}}_{\text{s}} - T^{\text{KS}}_{\text{s}}}
    {T^{\text{KS}}_{\text{s}}} ,
\end{equation}
and is plotted in Fig.~\ref{fig:de_contribs_pbex_h_atom}.
As can be seen, $r_{\text{kin}}$ becomes more and more
positive as we move from HF ($a = 1$; exact, no delocalisation error) 
to PBE ($a= 0$; delocalisation error), meaning that the kinetic energy 
obtained using the
density functional decreases compared to the HF kinetic energy. 
As reported by Mezei \textit{et al.}, HF can yield quite erroneous densities
but with good gradients and Laplacians.\cite{mezei2017} We therefore consider
another indicator, the scaled one-electron energy indicator, to avoid biasing
towards derivatives. The scaled one-electron energy indicator is given by
\begin{equation}
    r_{\text{s-1e}} = 
    \frac{E^{\text{HF}}_{\text{en}} - E^{\text{KS}}_{\text{en}}}
    {E^{\text{KS}}_{\text{en}}} -
    \frac{T^{\text{HF}}_{\text{s}} - T^{\text{KS}}_{\text{s}}}
    {T^{\text{KS}}_{\text{s}}}
\end{equation}
The idea behind the scaling is to put an equal weight on the density itself and
its derivatives. This time, the calculation is considered abnormal 
if $r_{\text{s-1e}}$ becomes negative.
As can be seen, the scaled one-electron indicator leads to the same conclusions
for this simple example. 

The \textit{virial theorem} in KS-DFT is given by\cite{rodriguez2009, levy1985} 
\begin{equation}
    \gamma [n] = \frac{V^{\text{KS}} [n] - T_{\text{c}} [n]}
    {T^{\text{KS}}_{\text{s}} [n] + T_{\text{c}} [n]} = -2.0
    \label{eq:virial_theorem}
\end{equation}
In Fig.~\ref{fig:de_contribs_pbex_h_atom} we further show another quantity, 
which is the change in the \textit{virial ratio} given by
\begin{equation}
    r_{\text{virial}} = \lvert \gamma' [n^{\text{HF}}] + 2.0 \rvert
    - \lvert \gamma' [n^{\text{DFT}}] + 2.0 \rvert
    \label{eq:change_vr}
\end{equation}
with the virial ratio defined as
\begin{align}
    \gamma' [n] &= \frac{V^{\text{KS}} [n]}
    {T^{\text{KS}}_{\text{s}} [n]} \label{eq:virial_ratio}\\
    V^{\text{KS}} [n] &= E_{\text{en}} + E_{\text{J}} 
    + E_{\text{xc}} + E_{\text{nn}}
\end{align}
As can be seen in Eq.~\ref{eq:change_vr}, $r_{\text{virial}}$ is negative if 
the virial ratio 
(Eq.~\ref{eq:virial_ratio}) evaluated on the HF density is closer to the  
value of $-2.0$ than the virial ratio evaluated on the self-consistent density.
We should highlight a few things here: First, when comparing
Eq.~\ref{eq:virial_ratio} and Eq.~\ref{eq:virial_theorem}, it is clear that the
virial ratio $\gamma'$ should not be expected to be exactly $-2.0$ --- due to
the fact that the correlation part of the kinetic energy $T_{\text{c}}$ is
(should be) included in the exchange-correlation energy --- but it is usually
quite close. Second, as can be seen in Fig.~\ref{fig:de_contribs_pbex_h_atom}, 
$r_{\text{virial}}$ could, in principle, serve as an indicator. However,
since the virial ratio only holds in the complete basis set limit and for atoms 
or molecules in their equilibrium geometry, we have chosen not to pursue it. 
Additionally, by inspecting Eq.~\ref{eq:virial_theorem} it could be expected 
that the virial ratio gets closer to $-2.0$ with increasing 
$T^{\text{KS}}_{\text{s}}$ (and decreasing $T_{\text{c}}$).
Third, inspecting Eq.~\ref{eq:virial_theorem} or~\ref{eq:virial_ratio} and 
considering that
$V^{\text{KS}}$ includes, in contrast to the HF functional, an energy
contribution stemming from electron correlation, it could be assumed that
$T^{\text{KS}}_{\text{s}}$ should be larger than its HF counterpart. 
We note that this ``contraction effect of correlation'' was also reported by
Baerends and co-workers,\cite{gritsenko1996} who found that 
$T_{\text{s}}^{\text{KS}} > T_{\text{s}}^{\text{HF}}$ holds true for all
of their investigated cases.
In this context it should be noted that although the definition in terms of 
orbitals is identical, 
the HF and the KS non-interacting kinetic energies are 
different,\cite{goerling1995}
since the HF method minimises the expectation value of the Hamiltonian over all
Slater determinants while the KS Slater determinant can only be constructed
from orbitals stemming from a local multiplicative potential yielding the exact
density according to
\begin{equation}
    \rho (\mathbf{r}) = 
    \sum_{i} \lvert \phi^{\text{KS}}_{i} (\mathbf{r}) \rvert^{2}
\end{equation}
However, that difference was shown to be small and this is why it is neglected
in DC(HF)-DFT.\cite{nam2020}
It is also true that, contrary to the KS case, no universal
proof exists that the HF kinetic energy \textit{needs to be} smaller than (or
equal to) the exact (interacting) kinetic energy; or in other words, that
$T_{\text{c}}$ needs to be non-negative. However, a realistic counter example
has not been found.\cite{crisostomo2022} 

\subsection{Sanity checks on typical normal and abnormal calculations}
To test the kinetic energy indicator, we evaluated it for various 
DFT calculations on the different systems contained in the 
S22\cite{jurecka2006, marshall2011} and B30\cite{bauza2013, roza2014}
test sets, serving as examples for normal and abnormal
calculations, respectively.\cite{sim2022} 
Both test sets were developed to assess the accuracy of a method in calculating
non-covalent interaction energies between molecules and complexes,
with high-level coupled-cluster calculations serving as a reference. 
The well-established S22 test set was designed to represent non-covalent
interactions in biological molecules in a balanced way (hydrogen bonds, weak
dispersion bonds, and mixed scenarios).
On the other hand, the B30 test set contains non-covalent interactions that 
showed to be challenging for especially pure density functionals: 
halogen bonds, chalcogen bonds, and pnicogen bonds.\cite{bauza2013}
With ``systems'' we thus mean the various complexes/dimers plus the
respective sub-systems/monomers. 

We used
several functionals for our tests: the LDA\cite{bloch1929, dirac1930,
vosko1980} functional as an
example known for large delocalisation errors; the 
PBE\cite{perdew1996, perdew1997} and the SCAN\cite{brandenburg2016,
furness2022} 
functionals since
they are probably the most popular non-empirical functionals in use today and
SCAN additionally fulfills many exact constraints; and
the M06-L\cite{zhao2006, zhao2008} functional as an example for a 
highly empirical functional.

\begin{figure}
    \includegraphics{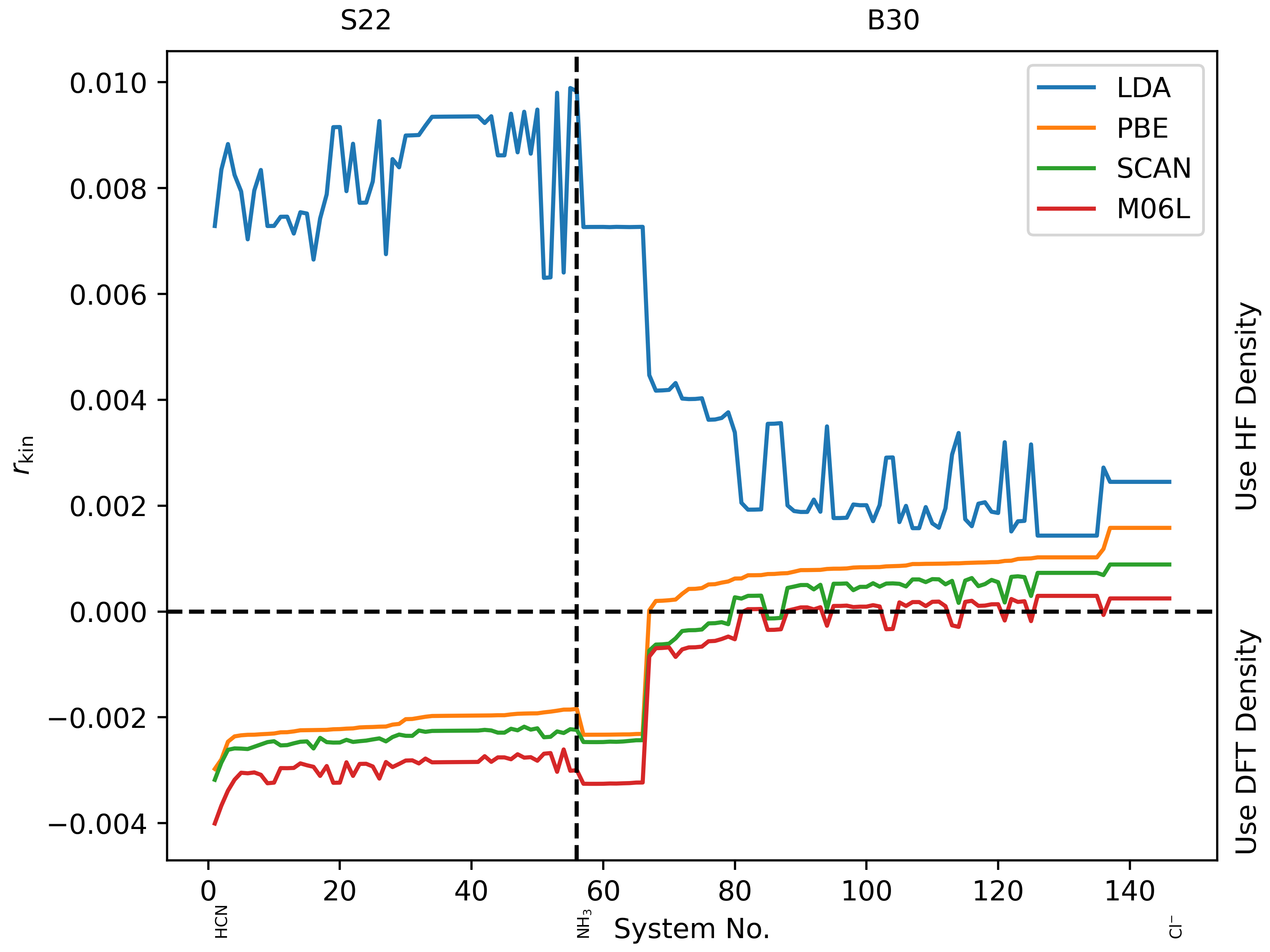}
    \caption{\label{fig:de_contribs_s22_b30_all_funcs} Relative change of the
    kinetic energy ($r_{\text{kin}}$) for different DFT calculations on the 
    S22 and B30 test sets.}
\end{figure}
Fig.~\ref{fig:de_contribs_s22_b30_all_funcs} shows $r_{\text{kin}}$ as
defined above for the different functionals and systems in the S22 and B30 test
sets. Note that all systems of a specific test set in this section were ordered 
by increasing value of $r_{\text{kin}}$ obtained with the PBE functional;
the complete ordered lists can be found in the supporting information.
As can be seen, for LDA the indicator is always larger than $0$ and hence 
\textit{always} suggests the use of the HF density. For the other three
functionals (PBE, SCAN, and M06-L) all calculations in the S22 test set are
predicted to be normal, whereas most of the calculations contained in the 
B30 test set are predicted to be abnormal. 
Since we chose the B30 test set to represent abnormal 
DFT calculations, these observations coincide exactly with our expectations.
Also note how the indicator changes for different functionals: based on this
indicator, the LDA density performs worse, followed by PBE, SCAN, and finally
M06-L. Furthermore, the three (m)GGAs seem to produce quite similar densities
according to our indicator. 
While this is interesting to observe, we stress that our
indicator is not intended to assess the quality of the different densities but
only to predict if the HF density is a better choice for a specific
calculation.

\begin{figure}
    \includegraphics{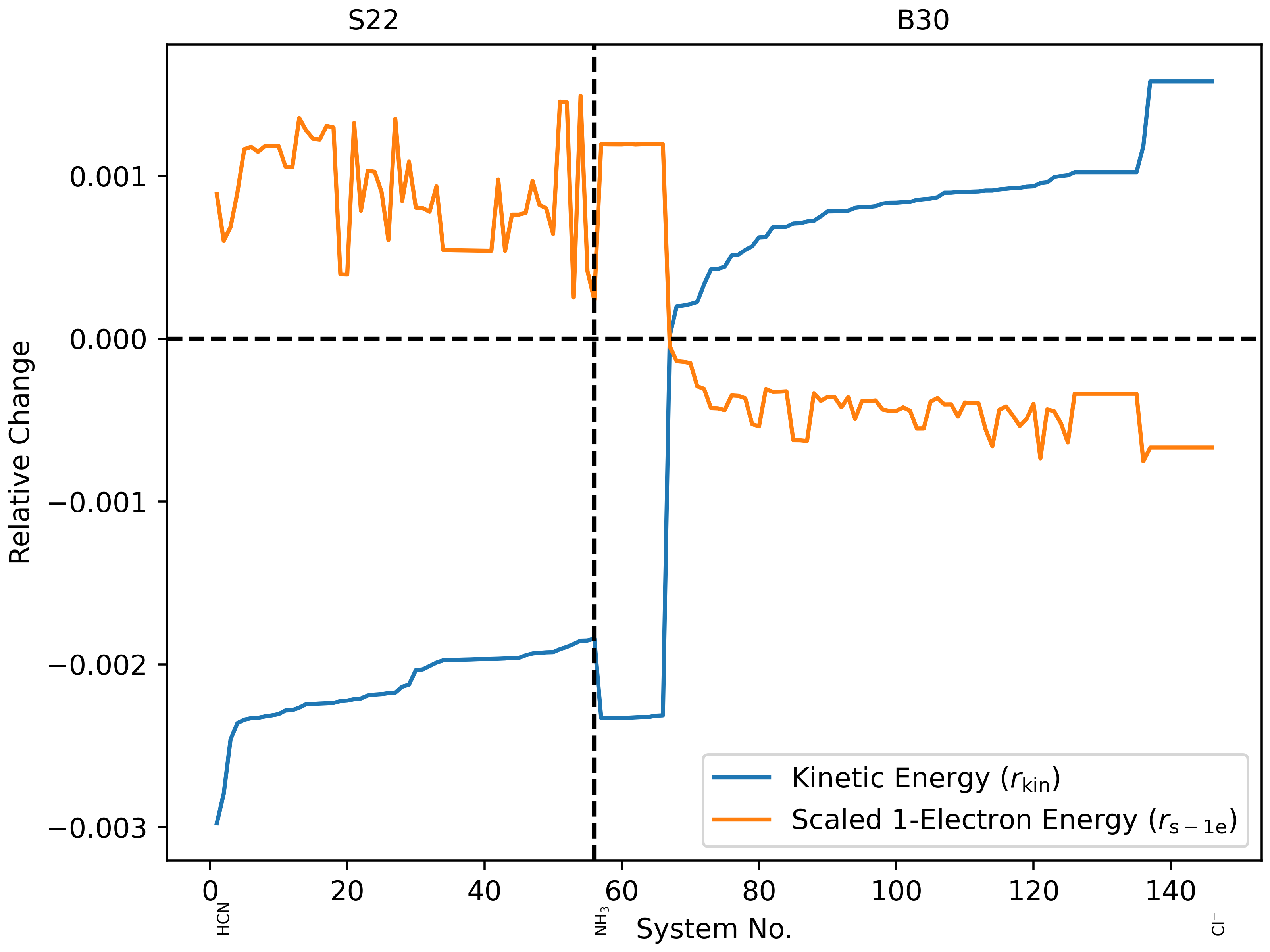}
    \caption{\label{fig:de_contribs_s22_b30_kin_s1e_pbe} Relative change of the
    kinetic energy ($r_{\text{kin}}$) and the scaled one-electron energy 
    ($r_{\text{s-1e}}$) for the PBE functional
    on the S22 and B30 test sets.}
\end{figure}
In Fig.~\ref{fig:de_contribs_s22_b30_kin_s1e_pbe} we additionally show 
$r_{\text{kin}}$ together with $r_{\text{s-1e}}$ for the PBE functional. 
As can be seen, as for
the H atom, the kinetic energy indicator and the scaled one-electron indicator
lead to the same conclusions (recall that abnormal calculations lead to
negative values for $r_{\text{s-1e}}$) 
and hence we will only use the kinetic energy
indicator in the following discussions.

\begin{figure}
    \includegraphics{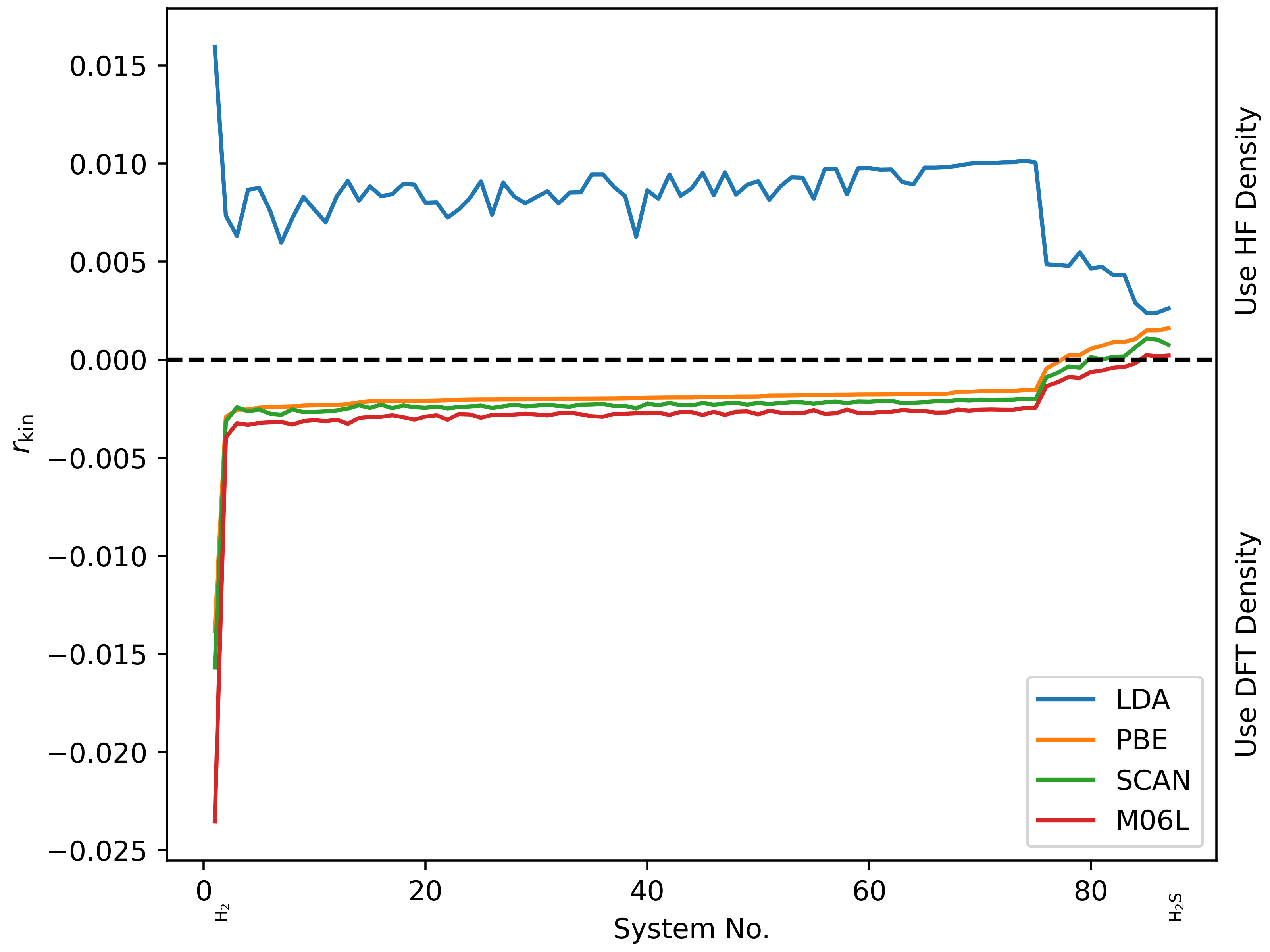}
    \caption{\label{fig:de_contribs_fh51_all_funcs} Relative change of the
    kinetic energy ($r_{\text{kin}}$) for different DFT calculations on the 
    FH51 test set. 
    Due to convergence problems for some systems, the
    cc-pVQZ\cite{dunning1989, woon1994, prascher2011, woon1993} 
    basis set was used for the SCAN functional.}
\end{figure}
We performed further sanity checks on a test set we would expect to include 
mostly normal calculations: 
the FH51 test set\cite{friedrich2013} consisting of reaction energies in small 
inorganic and organic systems. 
The results are shown in Fig.~\ref{fig:de_contribs_fh51_all_funcs}. 

As can be seen, the results
are in line with our expectations: in case of LDA the HF density should always
be the better choice, while the densities produced by the other three
functionals should be perfectly normal in the vast majority of cases.

We mentioned before that one of the characteristics an indicator should have in
our opinion is to avoid error cancellation as much as possible. The two main
sources of error cancellation as mentioned above are the use of an approximate
exchange-correlation functional to decide whether the HF density should be
used or not and further considering whole reactions instead of single 
calculations.
Although, of course, there is the virial ratio connecting $V^{\text{KS}}$ and
$T_{\text{s}}$, we are convinced that using the (non-interacting) 
kinetic energy functional --- which is exactly given in terms of orbitals --- 
is a step in the right direction
when it comes to avoiding the first source of error cancellation.
Addressing the second source of error cancellation is simple: we consider a
reaction abnormal --- and hence perform all necessary calculations using the HF
density --- if \textit{one} of the calculations is abnormal.

In the following, we will assess how well this procedure works by benchmarking
the accuracies of the resulting DC(HF)-DFT methods for different test sets 
taken from the GMTKN55 database.\cite{goerigk2017}

\subsection{Performance}

Let us start with the performance for the non-covalent interaction energies 
contained in the S22 and the B30 test sets. 
In the last section, it was shown that the kinetic energy indicator always 
suggests the use of the HF density for LDA. 
As can be seen in
Table~\ref{tbl:maes_all_functionals}, this leads to a significant lowering of
the mean absolute error (MAE) for both test sets. That the HF density performs
better than the LDA density is in line with observations presented in related
works.\cite{mostafanejad2019, mezei2017} 
For the other three
functionals the conclusion is the same: the kinetic energy indicator suggests
the ``more accurate'' density in both cases; the self-consistent one for the 
S22 and the HF one for the B30 test set. 

We further tested our kinetic energy indicator for the chemical problems 
included in 
the FH51, the G21EA\cite{curtiss1991, goerigk2010, goerigk2017} (adiabatic
electron affinities), and the 
DARC\cite{goerigk2010, goerigk2017, johnson2008} (Diels-Alder reactions) 
test sets. 
Overall, the kinetic energy indicator behaves as desired and leads
to significant improvements when the DFT densities are erroneous.
\begin{table}
  \caption{Mean absolute errors in kcal/mol of different functionals for 
    different test sets.}
  \label{tbl:maes_all_functionals}
  \begin{tabular}{ccccccccc}
    \hline
       & S22  & B30 & FH51\textsuperscript{\emph{a}} & G21EA & DARC \\
    \hline
      LDA & 2.18 & 8.26 & 6.69 & 7.83 & 11.83 \\
      LDA@HF & 1.35 & 5.02 & 5.44 & 6.92 & 8.86 \\
      DC(HF)-LDA & 1.35 & 5.02 & 5.44 & 6.73 & 8.86 \\
    \hline
      PBE & 2.56 & 2.46 & 3.44 & 3.69 & 6.63 \\
      PBE@HF & 3.26 & 1.00 & 3.44 & 2.93 & 7.64 \\
      DC(HF)-PBE & 2.56 & 1.00 & 3.55 & 3.04 & 6.63 \\
    \hline
      SCAN & 1.16 & 2.48 & 2.99 & 3.37 & 2.89 \\
      SCAN@HF & 1.54 & 0.77 & 2.56 & 4.20 & 3.41 \\
      DC(HF)-SCAN & 1.16 & 0.78 & 2.96 & 3.29 & 2.89 \\
    \hline
      M06-L & 0.72 & 1.34 & 2.84 & 3.46 & 8.15 \\
      M06-L@HF & 0.85 & 0.77 & 2.07 & 4.13 & 5.39 \\
      DC(HF)-M06-L & 0.72 & 0.86 & 2.81 & 3.48 & 8.15 \\
    \hline
  \end{tabular}

    {\tiny \textsuperscript{\emph{a}} Due to convergence problems for some 
    systems, the
    cc-pVQZ\cite{dunning1989, woon1994, prascher2011, woon1993} 
    basis set was used for the SCAN functional.}
\end{table}
However, questions about the reliability of our indicator arise when evaluating
the DARC test set using the \mbox{M06-L} functional.
In this case, the kinetic energy indicator clearly favours the ``less
accurate'' density. We conducted further examination to understand this
behaviour better.

\begin{figure}
    \includegraphics{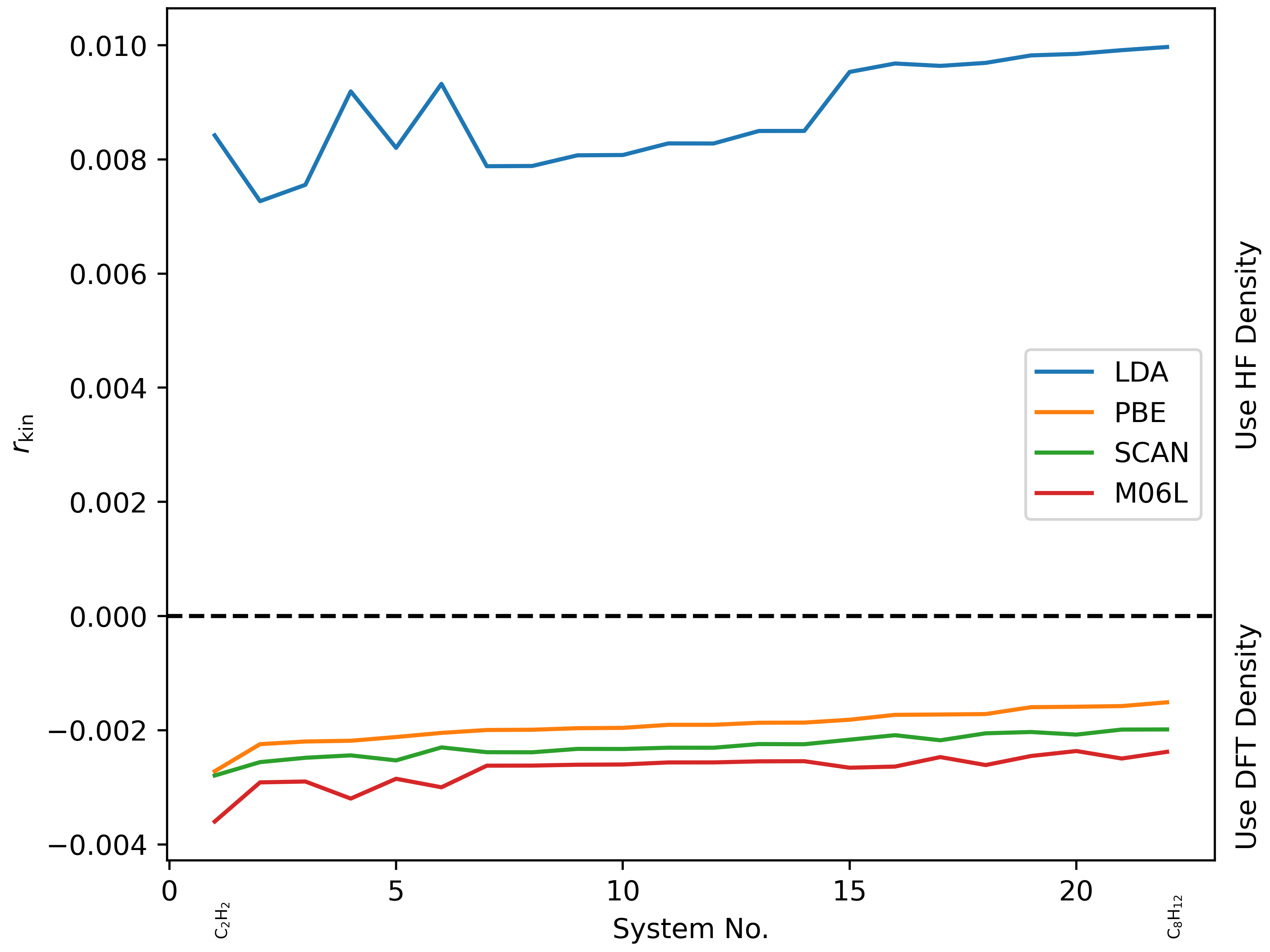}
    \caption{\label{fig:de_contribs_darc_all_funcs} Relative change of the
    kinetic energy ($r_{\text{kin}}$) for different DFT calculations on the 
    DARC test set.}
\end{figure}
\begin{figure}
    \includegraphics{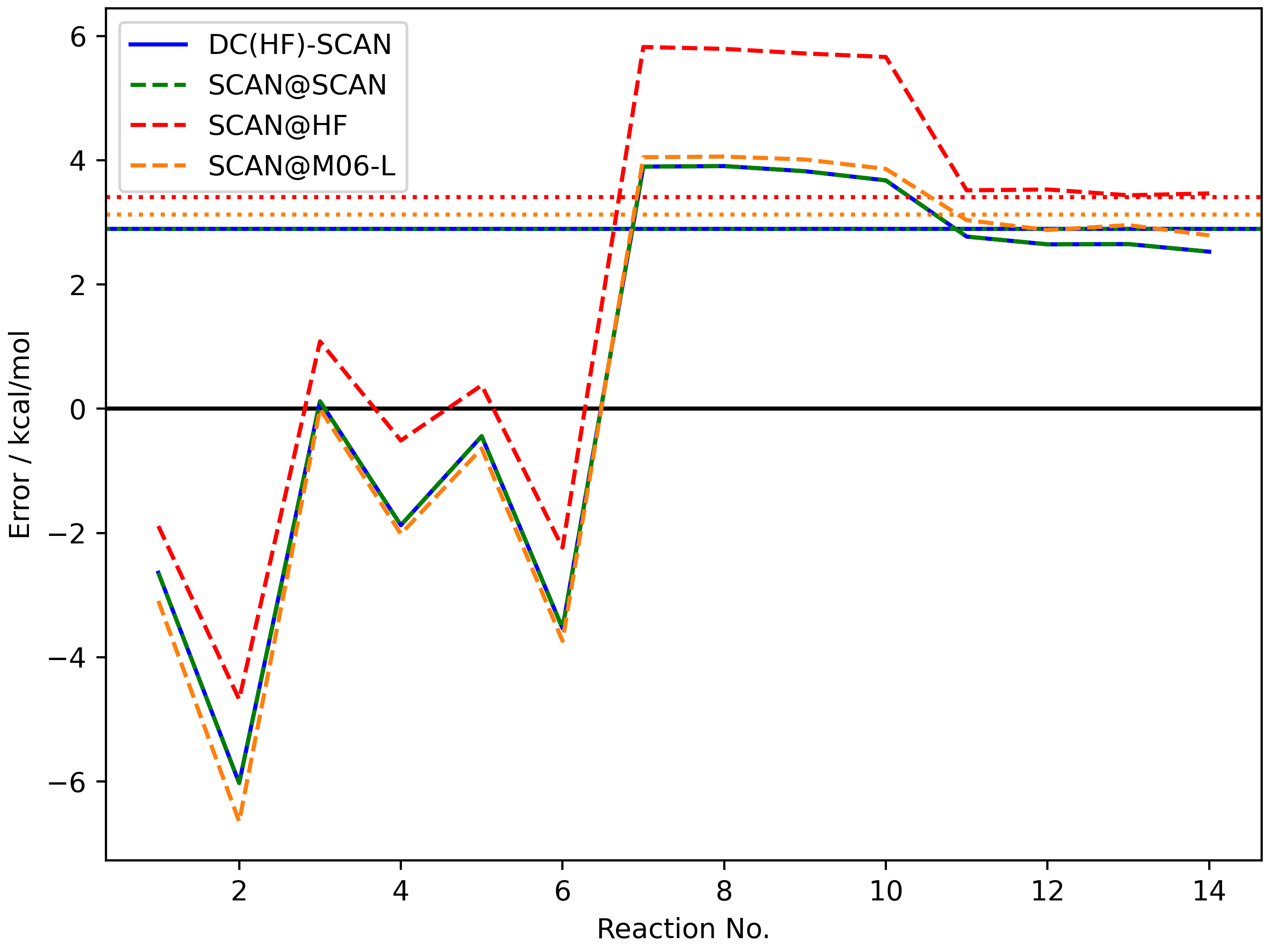}
    \caption{\label{fig:errors_scan_m06_darc} Errors in kcal/mol for the 
    different reactions contained in the DARC test set using the SCAN 
    functional on different densities. The coloured horizontal lines show the 
    respective mean absolute errors.}
\end{figure}

Fig.~\ref{fig:de_contribs_darc_all_funcs} shows $r_{\text{kin}}$ for the
calculations in the DARC test set performed 
with the LDA, PBE, SCAN, and M06-L functionals. 
As can be seen, only for
the LDA functional the kinetic energy indicator suggests the use of the HF
density. Furthermore, as for the examples presented in the last section, the
densities produced by the different (m)GGAs seem to be quite similar (at least
according to our kinetic energy indicator). Also note that the kinetic energy
indicator performs very well for all functionals except M06-L. Therefore, we
tested how the SCAN functional --- performing best on the DARC test set ---
performs when evaluated on the M06-L density. The results are shown in
Fig.~\ref{fig:errors_scan_m06_darc}. 

As can be seen, the errors of the SCAN functional evaluated on its 
self-consistent density and on the M06-L density are indeed very similar 
and hence the kinetic energy indicator correctly predicts the M06-L density 
to be normal --- if the SCAN density is normal then the M06-L density should be
normal as well. 
Therefore, the better performance of M06-L@HF is probably due to a fortuitous 
cancellation of the functional error and the errors in the HF density. Although
this behaviour of our kinetic energy indicator leads to worse results in this
case, it is still encouraging that it is able to make this distinction. We
assume similar reasons for the slight worsening of the PBE results for the 
FH51 test set.

\subsection{Efficiency}

As mentioned before, another key feature a good indicator should possess is
efficiency. So far, our kinetic energy indicator does not seem to improve a lot
upon
the density sensitivity put forward by Burke and co-workers in this respect. 
In order to address this, we propose the following procedure:

First, converge the DFT calculation. Second, use the converged DFT density as
initial guess for a HF calculation. Third, evaluate one Fock matrix and update
the orbitals and density. Fourth, evaluate the kinetic energy using the updated
orbitals and compare it with the converged DFT kinetic energy. Fifth, only
converge the HF calculation if 
$T_{\text{s}}^{\text{HF, 1-iter}} > T_{\text{s}}^{\text{KS}}$.

\begin{figure}
    \includegraphics{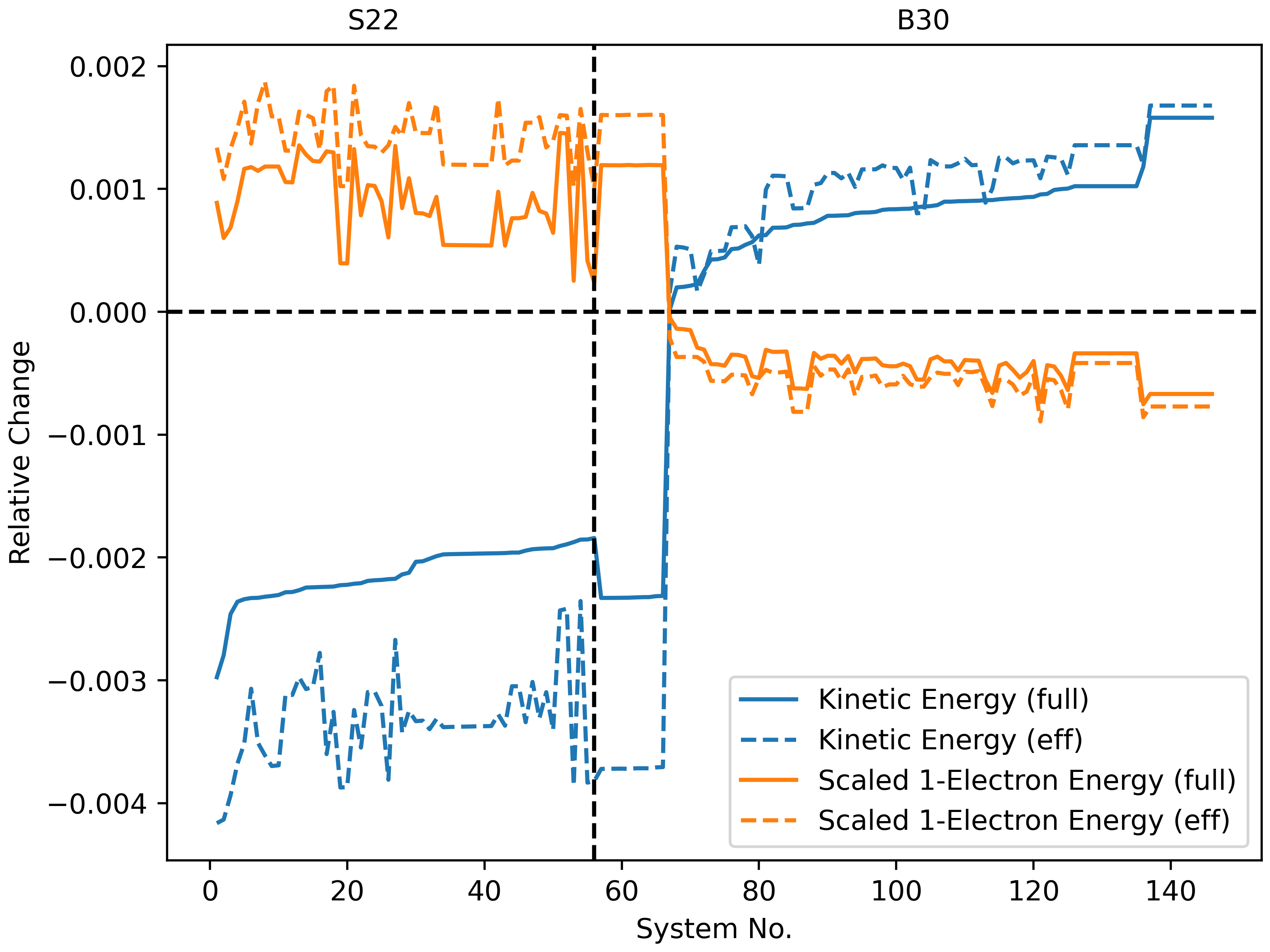}
    \caption{\label{fig:de_contribs_comp_s22_b30} Comparison of the two
    indicators evaluated with unconverged HF orbitals (only one HF iteration;
    eff) with their converged counterparts (full).}
\end{figure}
We investigated that scheme for the S22 and the B30 test sets using the PBE
functional. The indicators $r_{\text{kin}}$ and $r_{\text{s-1e}}$ after
only one HF iteration (denoted with ``eff'') and the converged 
counterparts (denoted with ``full'') are shown in
Fig.~\ref{fig:de_contribs_comp_s22_b30}. As can be seen, the indicators after 
only one HF iteration lead to the same results. Moreover, the unconverged
indicators tend to be larger in magnitude, which is ideal since it
ensures correct predictions.

\begin{figure}
    \includegraphics{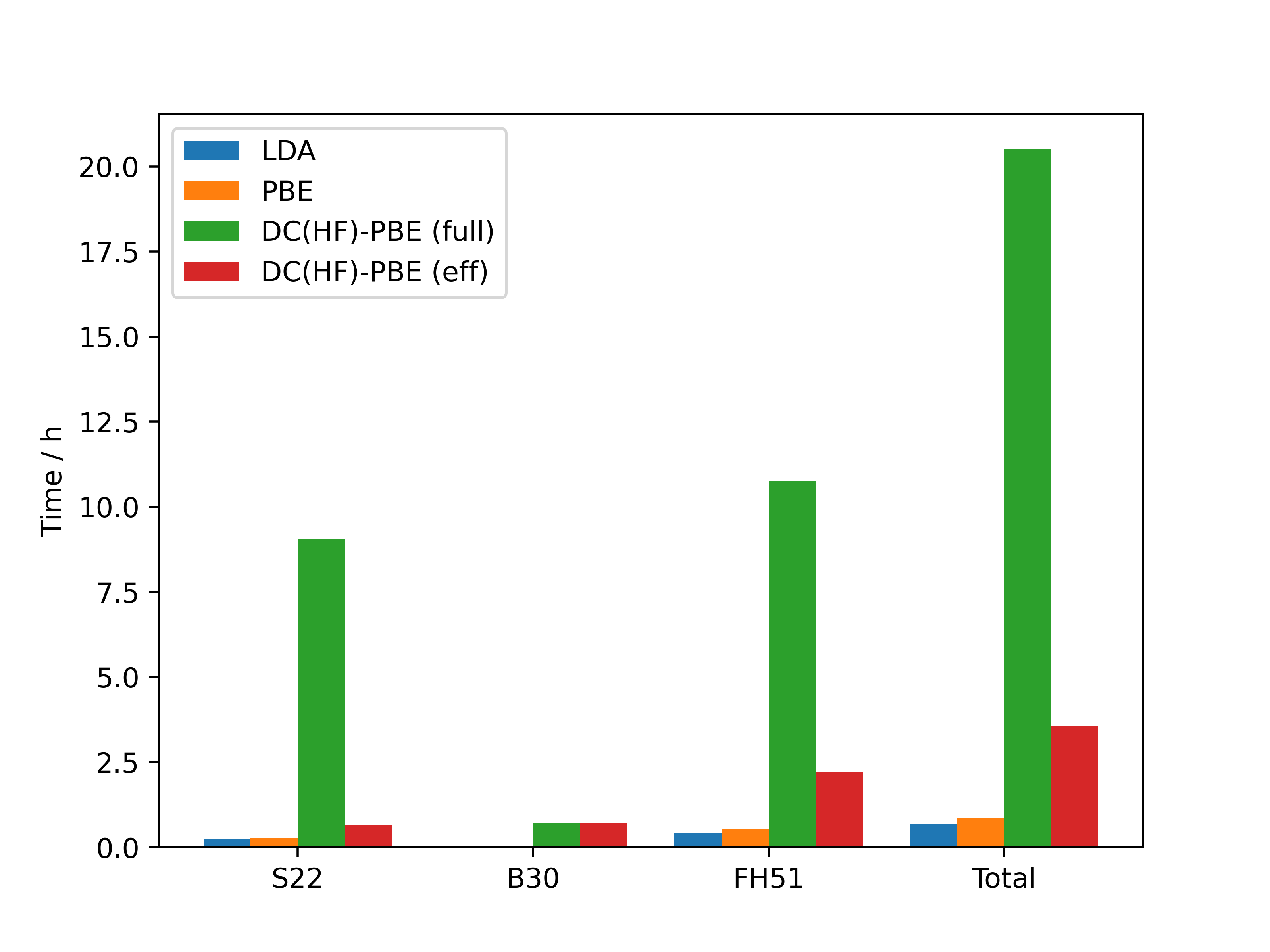}
    \caption{\label{fig:timings_s22_b30_fh51} Cumulative timings for the 
    S22, B30, and FH51 test sets.}
\end{figure}
Fig.~\ref{fig:timings_s22_b30_fh51} shows cumulative timings of pure LDA and
PBE, as well as full DC(HF)-PBE (converging the HF
calculation to assess whether the HF or the self-consistent density should be
used) and efficient DC(HF)-PBE 
(only one iteration of HF for the assessment) for the S22, B30,
and FH51 test sets; additionally, the time needed for all test sets together is
shown. The reason why we also show LDA timings is the fact that the LDA as well
the HF density are needed to evaluate the density sensitivity according to
Eq.~\ref{eq:dens_sens}.

To start with, we note that the HF calculations are significantly 
more expensive than both the LDA and the PBE calculations; 
in fact, the difference between LDA and PBE is negligible. 
Second, the savings in terms of computational cost are enormous when 
our efficient DC(HF)-PBE method is employed, and, of course, get even larger
the more normal calculations are included. We want to stress again that the
vast majority of DFT calculations is normal and hence our proposed procedure is
an important step to make the use of DC(HF)-DFT more routine.

Finally, although it was not necessary in the cases 
investigated here, it should be noted that, if the efficient indicator suggests
the use of the HF density, it is, of course, possible and probably also 
advisable to
check the indicator again after the HF calculation is fully converged; there is
no disadvantage in doing that. Additionally, the density sensitivity could be
evaluated with a small extra cost to introduce a further control mechanism. 
In that way, the density sensitivity and the
kinetic energy indicator can be considered complementary.

\subsection{Beyond density corrections}
In the last section, we proposed a scheme that significantly improves the
efficiency of our kinetic energy indicator. In this section, we want to go one
step further: since our indicator necessitates one iteration of HF in any case,
it naturally lends itself to including exact (HF) exchange in the final energy
and, in that way, additionally ``correcting'' the \textit{functional}. 
Consider the PBE functional as an example:

First, we converge a PBE calculation. After that, we use the PBE density to
evaluate the Fock matrix, update the orbitals, and compare the updated kinetic
energy with the one obtained using the PBE functional. If the PBE kinetic
energy is larger, we already have everything we need to evaluate the PBE0
functional on the PBE density. If the updated kinetic energy is larger, we
converge the HF calculation and the only thing we need to do now is to evaluate
the PBE exchange-correlation potential using the HF density on top of that,
which is, as can be seen in Fig.~\ref{fig:timings_s22_b30_fh51}, 
almost negligible in terms of computational cost. We note that the
choice which hybrid to evaluate in this step is completely flexible.
Instead of DC(HF)-DFT we call this procedure C(HF)-DFT (``corrected'' instead
of ``density corrected''), or for the specific case of PBE, C(HF)-PBE.

\begin{table}
  \caption{Mean absolute errors in kcal/mol of different functionals for 
    different test sets.}
  \label{tbl:maes_pbe_hybrid}
  \begin{tabular}{ccccccccc}
    \hline
       & S22  & B30 & FH51 & G21EA & DARC \\
    \hline
      PBE & 2.56 & 2.46 & 3.44 & 3.69 & 6.63 \\
      DC(HF)-PBE & 2.56 & 1.00 & 3.55 & 3.04 & 6.63 \\
      C(HF)-PBE & 2.41 & 0.90 & 2.62 & 2.59 & 3.08 \\
      PBE0 & 2.38 & 1.62 & 2.63 & 2.53 & 3.05 \\
    \hline
  \end{tabular}
\end{table}
We tested the proposed method on the test sets already used above. The results
are shown in Table~\ref{tbl:maes_pbe_hybrid}. 
As can be seen, C(HF)-PBE significantly improves upon pure PBE and performs
similarly to full (pure) PBE0. 
It is also worth noting the improvement of
C(HF)-PBE compared to PBE0 for the B30 test set, which is due to the use of the
HF density in that case.

\section{Computational details}
All calculations were carried out using a development version of the FermiONs++
programme package developed in the Ochsenfeld group.\cite{kussmann2013,
kussmann2015, kussmann2017} The binary has been compiled with the GNU Compiler
Collection (GCC) version 12.1. The calculations were executed on a compute
node containing 2 Intel Xeon E5-2630 v4 CPUs (20 cores / 40 threads;
2.20\,GHz). 
All runtimes given are wall times, not CPU times.

The evaluations of the exchange-correlation terms were performed using the
multi-grids defined in Ref.~\citenum{laqua2018} 
(smaller grid within the SCF optimization 
and larger grid for the final energy evaluation), 
generated with the modified Becke weighting scheme.\cite{laqua2018}
The SCF convergence threshold was set to $10^{-6}$ for the norm of the
difference density matrix $|| \Delta \mathbf{P} ||$.

We employ the integral-direct resolution-of-the-identity Coulomb (RI-J) 
method of Kussmann \textit{et al.}\cite{kussmann2021} for the
evaluation of the Coulomb matrices and the linear-scaling semi-numerical exact
exchange (sn-LinK) method of Laqua \textit{et al.}\cite{laqua2020} 
for the evaluation of the exact exchange matrices.

For the calculations on the H atom, the
def2-QZVPPD\cite{weigend2003, weigend2005, rappoport2010} basis set 
together with the
def2-universal-JFIT\cite{weigend2006} auxiliary basis set was employed. 
If not stated otherwise,
all calculations included in the test sets S22, B30, FH51, G21EA, and DARC were
performed using the aug-cc-pVQZ\cite{dunning1989, kendall1992, woon1994,
prascher2011, woon1993} basis set in combination with the 
cc-pVTZ-JKFIT\cite{weigend2002} auxiliary basis set.

\section{Conclusion}
In conclusion, we presented a simple yet efficient procedure to perform
DC(HF)-DFT calculations.
In this procedure, the crucial step of deciding whether the
self-consistent or the HF density should be used to evaluate the density
functional is conducted employing a simple heuristic based on the difference
between
the non-interacting kinetic energies obtained from the analysed functional and
the HF method, called the kinetic energy indicator. 
Our kinetic energy indicator offers several key characteristics that make it
stand out from other methods: 
Firstly, it directly compares the self-consistent
density of the analysed functional with the HF density. Secondly, it is
size-intensive, meaning that it is suitable for use in both large and small
systems. Thirdly, it reduces the probability of error cancellation, making it
more reliable. Finally, it is highly efficient.
We further note that our kinetic energy indicator is extremely simple to apply
in a retrospective analysis of DFT calculations, assuming that the 
non-interacting kinetic energies of the analysed DFT calculations are known. 
All that is necessary is to converge a HF calculation and compare the two
non-interacting kinetic energies.

It was shown that the
kinetic energy indicator reliably detects calculations where the use of the HF
density leads to improved results. Furthermore, the high efficiency of our
indicator was demonstrated on three different test sets contained in the
GMTKN55 database.

In addition, we have introduced a new procedure, called C(HF)-DFT, 
which not only
corrects the density if necessary, but also ``corrects'' the functional by 
evaluating a related hybrid 
at almost no extra computational cost. 
We have demonstrated its effectiveness using the PBE
functional, showing a significant improvement in accuracy that is comparable to
that of its parent hybrid, PBE0. 
Additionally, if the parent hybrid suffers from a density-driven error,
C(HF)-DFT can achieve even higher accuracy. 
Extending this procedure to double-hybrids is work in progress. 

Overall, our presented methods provide simple and effective solutions for 
improving density functional evaluations. 
As Burke and co-workers have noted,\cite{crisostomo2022b} even small
improvements in our current density functional approximations can have a
significant impact on applications in science and technology. Therefore, we
hope that our contributions will lead to more widespread application of
DC(HF)-DFT and C(HF)-DFT, and, in that way, have a 
positive impact on quantum chemical applications of all kinds.

\begin{acknowledgement}
    D.~G.~acknowledges funding by the Deutsche Forschungsgemeinschaft 
    (DFG, German Research Foundation) -- 498448112. 
    D.~G.~thanks J.~Kussmann (LMU Munich) for providing a development version 
    of the FermiONs++ programme package.
\end{acknowledgement}

\begin{suppinfo}

The following files are available free of charge:
\begin{itemize}
  \item lists\_of\_energy\_contributions.xlsx: Excel file containing all
      relevant energy contributions for all systems and functionals considered
        in this work
  \item ordered\_lists\_of\_systems.txt: Text file containing the ordered lists
      of systems for the S22, B30, FH51, and DARC test sets
  \item geometries\_and\_references.zip: All geometries and reference values
      used in this work
\end{itemize}

\end{suppinfo}

\providecommand{\latin}[1]{#1}
\makeatletter
\providecommand{\doi}
  {\begingroup\let\do\@makeother\dospecials
  \catcode`\{=1 \catcode`\}=2 \doi@aux}
\providecommand{\doi@aux}[1]{\endgroup\texttt{#1}}
\makeatother
\providecommand*\mcitethebibliography{\thebibliography}
\csname @ifundefined\endcsname{endmcitethebibliography}
  {\let\endmcitethebibliography\endthebibliography}{}


\begin{mcitethebibliography}{110}
\providecommand*\natexlab[1]{#1}
\providecommand*\mciteSetBstSublistMode[1]{}
\providecommand*\mciteSetBstMaxWidthForm[2]{}
\providecommand*\mciteBstWouldAddEndPuncttrue
  {\def\EndOfBibitem{\unskip.}}
\providecommand*\mciteBstWouldAddEndPunctfalse
  {\let\EndOfBibitem\relax}
\providecommand*\mciteSetBstMidEndSepPunct[3]{}
\providecommand*\mciteSetBstSublistLabelBeginEnd[3]{}
\providecommand*\EndOfBibitem{}
\mciteSetBstSublistMode{f}
\mciteSetBstMaxWidthForm{subitem}{(\alph{mcitesubitemcount})}
\mciteSetBstSublistLabelBeginEnd
  {\mcitemaxwidthsubitemform\space}
  {\relax}
  {\relax}

\bibitem[Vuckovic \latin{et~al.}(2019)Vuckovic, Song, Kozlowski, Sim, and
  Burke]{vuckovic2019}
Vuckovic,~S.; Song,~S.; Kozlowski,~J.; Sim,~E.; Burke,~K. Density Functional
  Analysis: The Theory of Density-Corrected DFT. \emph{J. Chem. Theory Comput.}
  \textbf{2019}, \emph{15}, 6636--6646\relax
\mciteBstWouldAddEndPuncttrue
\mciteSetBstMidEndSepPunct{\mcitedefaultmidpunct}
{\mcitedefaultendpunct}{\mcitedefaultseppunct}\relax
\EndOfBibitem
\bibitem[Kohn and Sham(1965)Kohn, and Sham]{kohn1965}
Kohn,~W.; Sham,~L.~J. Self-Consistent Equations Including Exchange and
  Correlation Effects. \emph{Physical Review} \textbf{1965}, \emph{140},
  A1133--A1138\relax
\mciteBstWouldAddEndPuncttrue
\mciteSetBstMidEndSepPunct{\mcitedefaultmidpunct}
{\mcitedefaultendpunct}{\mcitedefaultseppunct}\relax
\EndOfBibitem
\bibitem[Mori-Sánchez \latin{et~al.}(2008)Mori-Sánchez, Cohen, and
  Yang]{sanchez2008}
Mori-Sánchez,~P.; Cohen,~A.~J.; Yang,~W. Localization and Delocalization
  Errors in Density Functional Theory and Implications for Band-Gap Prediction.
  \emph{Phys. Rev. Lett.} \textbf{2008}, \emph{100}, 146401\relax
\mciteBstWouldAddEndPuncttrue
\mciteSetBstMidEndSepPunct{\mcitedefaultmidpunct}
{\mcitedefaultendpunct}{\mcitedefaultseppunct}\relax
\EndOfBibitem
\bibitem[Cohen \latin{et~al.}(2007)Cohen, Mori-Sánchez, and Yang]{cohen2007}
Cohen,~A.~J.; Mori-Sánchez,~P.; Yang,~W. Development of exchange-correlation
  functionals with minimal many-electron self-interaction error. \emph{J. Chem.
  Phys.} \textbf{2007}, \emph{126}, 191109\relax
\mciteBstWouldAddEndPuncttrue
\mciteSetBstMidEndSepPunct{\mcitedefaultmidpunct}
{\mcitedefaultendpunct}{\mcitedefaultseppunct}\relax
\EndOfBibitem
\bibitem[Cohen \latin{et~al.}(2008)Cohen, Mori-Sánchez, and Yang]{cohen2008}
Cohen,~A.~J.; Mori-Sánchez,~P.; Yang,~W. Insights into Current Limitations of
  Density Functional Theory. \emph{Science} \textbf{2008}, \emph{321},
  792--794\relax
\mciteBstWouldAddEndPuncttrue
\mciteSetBstMidEndSepPunct{\mcitedefaultmidpunct}
{\mcitedefaultendpunct}{\mcitedefaultseppunct}\relax
\EndOfBibitem
\bibitem[Li \latin{et~al.}(2017)Li, Zheng, Su, and Yang]{li2017}
Li,~C.; Zheng,~X.; Su,~N.~Q.; Yang,~W. Localized orbital scaling correction for
  systematic elimination of delocalization error in density functional
  approximations. \emph{Natl. Sci. Rev.} \textbf{2017}, \emph{5},
  203--215\relax
\mciteBstWouldAddEndPuncttrue
\mciteSetBstMidEndSepPunct{\mcitedefaultmidpunct}
{\mcitedefaultendpunct}{\mcitedefaultseppunct}\relax
\EndOfBibitem
\bibitem[Li \latin{et~al.}(2015)Li, Zheng, Cohen, Mori-Sánchez, and
  Yang]{li2015}
Li,~C.; Zheng,~X.; Cohen,~A.~J.; Mori-Sánchez,~P.; Yang,~W. Local Scaling
  Correction for Reducing Delocalization Error in Density Functional
  Approximations. \emph{Phys. Rev. Lett.} \textbf{2015}, \emph{114},
  053001\relax
\mciteBstWouldAddEndPuncttrue
\mciteSetBstMidEndSepPunct{\mcitedefaultmidpunct}
{\mcitedefaultendpunct}{\mcitedefaultseppunct}\relax
\EndOfBibitem
\bibitem[Johnson \latin{et~al.}(2013)Johnson, Otero-de-la Roza, and
  Dale]{johnson2013}
Johnson,~E.~R.; Otero-de-la Roza,~A.; Dale,~S.~G. Extreme density-driven
  delocalization error for a model solvated-electron system. \emph{J. Chem.
  Phys.} \textbf{2013}, \emph{139}, 184116\relax
\mciteBstWouldAddEndPuncttrue
\mciteSetBstMidEndSepPunct{\mcitedefaultmidpunct}
{\mcitedefaultendpunct}{\mcitedefaultseppunct}\relax
\EndOfBibitem
\bibitem[Vazquez and Isborn(2015)Vazquez, and Isborn]{vazquez2015}
Vazquez,~X. A.~S.; Isborn,~C.~M. Size-dependent error of the density functional
  theory ionization potential in vacuum and solution. \emph{J. Chem. Phys.}
  \textbf{2015}, \emph{143}, 244105\relax
\mciteBstWouldAddEndPuncttrue
\mciteSetBstMidEndSepPunct{\mcitedefaultmidpunct}
{\mcitedefaultendpunct}{\mcitedefaultseppunct}\relax
\EndOfBibitem
\bibitem[LeBlanc \latin{et~al.}(2018)LeBlanc, Dale, Taylor, Becke, Day, and
  Johnson]{leblanc2018}
LeBlanc,~L.~M.; Dale,~S.~G.; Taylor,~C.~R.; Becke,~A.~D.; Day,~G.~M.;
  Johnson,~E.~R. Pervasive Delocalisation Error Causes Spurious Proton Transfer
  in Organic Acid–Base Co-Crystals. \emph{Angew. Chem. Int. Ed.}
  \textbf{2018}, \emph{57}, 14906--14910\relax
\mciteBstWouldAddEndPuncttrue
\mciteSetBstMidEndSepPunct{\mcitedefaultmidpunct}
{\mcitedefaultendpunct}{\mcitedefaultseppunct}\relax
\EndOfBibitem
\bibitem[Perdew and Zunger(1981)Perdew, and Zunger]{perdew1981}
Perdew,~J.~P.; Zunger,~A. Self-interaction correction to density-functional
  approximations for many-electron systems. \emph{Phys. Rev. B} \textbf{1981},
  \emph{23}, 5048--5079\relax
\mciteBstWouldAddEndPuncttrue
\mciteSetBstMidEndSepPunct{\mcitedefaultmidpunct}
{\mcitedefaultendpunct}{\mcitedefaultseppunct}\relax
\EndOfBibitem
\bibitem[Mori-Sánchez \latin{et~al.}(2006)Mori-Sánchez, Cohen, and
  Yang]{sanchez2006}
Mori-Sánchez,~P.; Cohen,~A.~J.; Yang,~W. Many-electron self-interaction error
  in approximate density functionals. \emph{J. Chem. Phys.} \textbf{2006},
  \emph{125}, 201102\relax
\mciteBstWouldAddEndPuncttrue
\mciteSetBstMidEndSepPunct{\mcitedefaultmidpunct}
{\mcitedefaultendpunct}{\mcitedefaultseppunct}\relax
\EndOfBibitem
\bibitem[Vydrov \latin{et~al.}(2007)Vydrov, Scuseria, and Perdew]{vydrov2007}
Vydrov,~O.~A.; Scuseria,~G.~E.; Perdew,~J.~P. Tests of functionals for systems
  with fractional electron number. \emph{J. Chem. Phys.} \textbf{2007},
  \emph{126}, 154109\relax
\mciteBstWouldAddEndPuncttrue
\mciteSetBstMidEndSepPunct{\mcitedefaultmidpunct}
{\mcitedefaultendpunct}{\mcitedefaultseppunct}\relax
\EndOfBibitem
\bibitem[Ruzsinszky \latin{et~al.}(2007)Ruzsinszky, Perdew, Csonka, Vydrov, and
  Scuseria]{ruzsinszky2007}
Ruzsinszky,~A.; Perdew,~J.~P.; Csonka,~G.~I.; Vydrov,~O.~A.; Scuseria,~G.~E.
  Density functionals that are one- and two- are not always many-electron
  self-interaction-free, as shown for H2+, He2+, LiH+, and Ne2+. \emph{J. Chem.
  Phys.} \textbf{2007}, \emph{126}, 104102\relax
\mciteBstWouldAddEndPuncttrue
\mciteSetBstMidEndSepPunct{\mcitedefaultmidpunct}
{\mcitedefaultendpunct}{\mcitedefaultseppunct}\relax
\EndOfBibitem
\bibitem[Perdew \latin{et~al.}(1982)Perdew, Parr, Levy, and Balduz]{perdew1982}
Perdew,~J.~P.; Parr,~R.~G.; Levy,~M.; Balduz,~J.~L. Density-Functional Theory
  for Fractional Particle Number: Derivative Discontinuities of the Energy.
  \emph{Phys. Rev. Lett.} \textbf{1982}, \emph{49}, 1691--1694\relax
\mciteBstWouldAddEndPuncttrue
\mciteSetBstMidEndSepPunct{\mcitedefaultmidpunct}
{\mcitedefaultendpunct}{\mcitedefaultseppunct}\relax
\EndOfBibitem
\bibitem[Mori-Sánchez \latin{et~al.}(2009)Mori-Sánchez, Cohen, and
  Yang]{sanchez2009}
Mori-Sánchez,~P.; Cohen,~A.~J.; Yang,~W. Discontinuous Nature of the
  Exchange-Correlation Functional in Strongly Correlated Systems. \emph{Phys.
  Rev. Lett.} \textbf{2009}, \emph{102}, 066403\relax
\mciteBstWouldAddEndPuncttrue
\mciteSetBstMidEndSepPunct{\mcitedefaultmidpunct}
{\mcitedefaultendpunct}{\mcitedefaultseppunct}\relax
\EndOfBibitem
\bibitem[Yang \latin{et~al.}(2012)Yang, Cohen, and Mori-Sánchez]{yang2012}
Yang,~W.; Cohen,~A.~J.; Mori-Sánchez,~P. Derivative discontinuity, bandgap and
  lowest unoccupied molecular orbital in density functional theory. \emph{J.
  Chem. Phys.} \textbf{2012}, \emph{136}, 204111\relax
\mciteBstWouldAddEndPuncttrue
\mciteSetBstMidEndSepPunct{\mcitedefaultmidpunct}
{\mcitedefaultendpunct}{\mcitedefaultseppunct}\relax
\EndOfBibitem
\bibitem[Zhang and Yang(1998)Zhang, and Yang]{zhang1998}
Zhang,~Y.; Yang,~W. A challenge for density functionals: Self-interaction error
  increases for systems with a noninteger number of electrons. \emph{J. Chem.
  Phys.} \textbf{1998}, \emph{109}, 2604--2608\relax
\mciteBstWouldAddEndPuncttrue
\mciteSetBstMidEndSepPunct{\mcitedefaultmidpunct}
{\mcitedefaultendpunct}{\mcitedefaultseppunct}\relax
\EndOfBibitem
\bibitem[Ruzsinszky \latin{et~al.}(2006)Ruzsinszky, Perdew, Csonka, Vydrov, and
  Scuseria]{ruzsinszky2006}
Ruzsinszky,~A.; Perdew,~J.~P.; Csonka,~G.~I.; Vydrov,~O.~A.; Scuseria,~G.~E.
  Spurious fractional charge on dissociated atoms: Pervasive and resilient
  self-interaction error of common density functionals. \emph{J. Chem. Phys.}
  \textbf{2006}, \emph{125}, 194112\relax
\mciteBstWouldAddEndPuncttrue
\mciteSetBstMidEndSepPunct{\mcitedefaultmidpunct}
{\mcitedefaultendpunct}{\mcitedefaultseppunct}\relax
\EndOfBibitem
\bibitem[Cohen and Mori-Sánchez(2014)Cohen, and Mori-Sánchez]{cohen2014}
Cohen,~A.~J.; Mori-Sánchez,~P. Dramatic changes in electronic structure
  revealed by fractionally charged nuclei. \emph{J. Chem. Phys.} \textbf{2014},
  \emph{140}, 044110\relax
\mciteBstWouldAddEndPuncttrue
\mciteSetBstMidEndSepPunct{\mcitedefaultmidpunct}
{\mcitedefaultendpunct}{\mcitedefaultseppunct}\relax
\EndOfBibitem
\bibitem[Cohen \latin{et~al.}(2012)Cohen, Mori-Sánchez, and Yang]{cohen2012}
Cohen,~A.~J.; Mori-Sánchez,~P.; Yang,~W. Challenges for Density Functional
  Theory. \emph{Chem. Rev.} \textbf{2012}, \emph{112}, 289--320\relax
\mciteBstWouldAddEndPuncttrue
\mciteSetBstMidEndSepPunct{\mcitedefaultmidpunct}
{\mcitedefaultendpunct}{\mcitedefaultseppunct}\relax
\EndOfBibitem
\bibitem[Hartree(1928)]{hartree1928}
Hartree,~D.~R. The Wave Mechanics of an Atom with a Non-Coulomb Central Field.
  Part I. Theory and Methods. \emph{Math. Proc. Cambridge Philos.}
  \textbf{1928}, \emph{24}, 89--110\relax
\mciteBstWouldAddEndPuncttrue
\mciteSetBstMidEndSepPunct{\mcitedefaultmidpunct}
{\mcitedefaultendpunct}{\mcitedefaultseppunct}\relax
\EndOfBibitem
\bibitem[Slater(1930)]{slater1930}
Slater,~J.~C. Note on Hartree's Method. \emph{Physical Review} \textbf{1930},
  \emph{35}, 210--211\relax
\mciteBstWouldAddEndPuncttrue
\mciteSetBstMidEndSepPunct{\mcitedefaultmidpunct}
{\mcitedefaultendpunct}{\mcitedefaultseppunct}\relax
\EndOfBibitem
\bibitem[Fock(1930)]{fock1930}
Fock,~V. Näherungsmethode zur Lösung des quantenmechanischen
  Mehrkörperproblems. \emph{Z. Physik} \textbf{1930}, \emph{61},
  126--148\relax
\mciteBstWouldAddEndPuncttrue
\mciteSetBstMidEndSepPunct{\mcitedefaultmidpunct}
{\mcitedefaultendpunct}{\mcitedefaultseppunct}\relax
\EndOfBibitem
\bibitem[Leininger \latin{et~al.}(1997)Leininger, Stoll, Werner, and
  Savin]{leininger1997}
Leininger,~T.; Stoll,~H.; Werner,~H.-J.; Savin,~A. Combining long-range
  configuration interaction with short-range density functionals. \emph{Chem.
  Phys. Lett.} \textbf{1997}, \emph{275}, 151--160\relax
\mciteBstWouldAddEndPuncttrue
\mciteSetBstMidEndSepPunct{\mcitedefaultmidpunct}
{\mcitedefaultendpunct}{\mcitedefaultseppunct}\relax
\EndOfBibitem
\bibitem[Gordon and Kim(1972)Gordon, and Kim]{gordon1972}
Gordon,~R.~G.; Kim,~Y.~S. Theory for the Forces between Closed‐Shell Atoms
  and Molecules. \emph{J. Chem. Phys.} \textbf{1972}, \emph{56},
  3122--3133\relax
\mciteBstWouldAddEndPuncttrue
\mciteSetBstMidEndSepPunct{\mcitedefaultmidpunct}
{\mcitedefaultendpunct}{\mcitedefaultseppunct}\relax
\EndOfBibitem
\bibitem[Colle and Salvetti(1975)Colle, and Salvetti]{colle1975}
Colle,~R.; Salvetti,~O. Approximate calculation of the correlation energy for
  the closed shells. \emph{Theoret. Chim. Acta} \textbf{1975}, \emph{37},
  329--334\relax
\mciteBstWouldAddEndPuncttrue
\mciteSetBstMidEndSepPunct{\mcitedefaultmidpunct}
{\mcitedefaultendpunct}{\mcitedefaultseppunct}\relax
\EndOfBibitem
\bibitem[Scuseria(1992)]{scuseria1992}
Scuseria,~G.~E. Comparison of coupled‐cluster results with a hybrid of
  Hartree–Fock and density functional theory. \emph{J. Chem. Phys.}
  \textbf{1992}, \emph{97}, 7528--7530\relax
\mciteBstWouldAddEndPuncttrue
\mciteSetBstMidEndSepPunct{\mcitedefaultmidpunct}
{\mcitedefaultendpunct}{\mcitedefaultseppunct}\relax
\EndOfBibitem
\bibitem[Janesko and Scuseria(2008)Janesko, and Scuseria]{janesko2008}
Janesko,~B.~G.; Scuseria,~G.~E. Hartree–Fock orbitals significantly improve
  the reaction barrier heights predicted by semilocal density functionals.
  \emph{J. Chem. Phys.} \textbf{2008}, \emph{128}, 244112\relax
\mciteBstWouldAddEndPuncttrue
\mciteSetBstMidEndSepPunct{\mcitedefaultmidpunct}
{\mcitedefaultendpunct}{\mcitedefaultseppunct}\relax
\EndOfBibitem
\bibitem[Cioslowski and Nanayakkara(1993)Cioslowski, and
  Nanayakkara]{cioslowski1993}
Cioslowski,~J.; Nanayakkara,~A. Electron correlation contributions to
  one‐electron properties from functionals of the Hartree–Fock electron
  density. \emph{J. Chem. Phys.} \textbf{1993}, \emph{99}, 5163--5166\relax
\mciteBstWouldAddEndPuncttrue
\mciteSetBstMidEndSepPunct{\mcitedefaultmidpunct}
{\mcitedefaultendpunct}{\mcitedefaultseppunct}\relax
\EndOfBibitem
\bibitem[Oliphant and Bartlett(1994)Oliphant, and Bartlett]{oliphant1994}
Oliphant,~N.; Bartlett,~R.~J. A systematic comparison of molecular properties
  obtained using Hartree–Fock, a hybrid Hartree–Fock
  density‐functional‐theory, and coupled‐cluster methods. \emph{J. Chem.
  Phys.} \textbf{1994}, \emph{100}, 6550--6561\relax
\mciteBstWouldAddEndPuncttrue
\mciteSetBstMidEndSepPunct{\mcitedefaultmidpunct}
{\mcitedefaultendpunct}{\mcitedefaultseppunct}\relax
\EndOfBibitem
\bibitem[Verma \latin{et~al.}(2012)Verma, Perera, and Bartlett]{verma2012}
Verma,~P.; Perera,~A.; Bartlett,~R.~J. Increasing the applicability of DFT I:
  Non-variational correlation corrections from Hartree–Fock DFT for
  predicting transition states. \emph{Chem. Phys. Lett.} \textbf{2012},
  \emph{524}, 10--15\relax
\mciteBstWouldAddEndPuncttrue
\mciteSetBstMidEndSepPunct{\mcitedefaultmidpunct}
{\mcitedefaultendpunct}{\mcitedefaultseppunct}\relax
\EndOfBibitem
\bibitem[Santra and Martin(2021)Santra, and Martin]{santra2021}
Santra,~G.; Martin,~J. M.~L. What Types of Chemical Problems Benefit from
  Density-Corrected DFT? A Probe Using an Extensive and Chemically Diverse Test
  Suite. \emph{J. Chem. Theory Comput.} \textbf{2021}, \emph{17},
  1368--1379\relax
\mciteBstWouldAddEndPuncttrue
\mciteSetBstMidEndSepPunct{\mcitedefaultmidpunct}
{\mcitedefaultendpunct}{\mcitedefaultseppunct}\relax
\EndOfBibitem
\bibitem[Nam \latin{et~al.}(2021)Nam, Cho, Sim, and Burke]{nam2021}
Nam,~S.; Cho,~E.; Sim,~E.; Burke,~K. Explaining and Fixing DFT Failures for
  Torsional Barriers. \emph{J. Phys. Chem. Lett.} \textbf{2021}, \emph{12},
  2796--2804\relax
\mciteBstWouldAddEndPuncttrue
\mciteSetBstMidEndSepPunct{\mcitedefaultmidpunct}
{\mcitedefaultendpunct}{\mcitedefaultseppunct}\relax
\EndOfBibitem
\bibitem[Nam \latin{et~al.}(2020)Nam, Song, Sim, and Burke]{nam2020}
Nam,~S.; Song,~S.; Sim,~E.; Burke,~K. Measuring Density-Driven Errors Using
  Kohn–Sham Inversion. \emph{J. Chem. Theory Comput.} \textbf{2020},
  \emph{16}, 5014--5023\relax
\mciteBstWouldAddEndPuncttrue
\mciteSetBstMidEndSepPunct{\mcitedefaultmidpunct}
{\mcitedefaultendpunct}{\mcitedefaultseppunct}\relax
\EndOfBibitem
\bibitem[Sim \latin{et~al.}(2018)Sim, Song, and Burke]{sim2018}
Sim,~E.; Song,~S.; Burke,~K. Quantifying Density Errors in DFT. \emph{J. Phys.
  Chem. Lett.} \textbf{2018}, \emph{9}, 6385--6392\relax
\mciteBstWouldAddEndPuncttrue
\mciteSetBstMidEndSepPunct{\mcitedefaultmidpunct}
{\mcitedefaultendpunct}{\mcitedefaultseppunct}\relax
\EndOfBibitem
\bibitem[Sim \latin{et~al.}(2022)Sim, Song, Vuckovic, and Burke]{sim2022}
Sim,~E.; Song,~S.; Vuckovic,~S.; Burke,~K. Improving Results by Improving
  Densities: Density-Corrected Density Functional Theory. \emph{J. Am. Chem.
  Soc.} \textbf{2022}, \emph{144}, 6625--6639\relax
\mciteBstWouldAddEndPuncttrue
\mciteSetBstMidEndSepPunct{\mcitedefaultmidpunct}
{\mcitedefaultendpunct}{\mcitedefaultseppunct}\relax
\EndOfBibitem
\bibitem[Song \latin{et~al.}(2021)Song, Vuckovic, Sim, and Burke]{song2021}
Song,~S.; Vuckovic,~S.; Sim,~E.; Burke,~K. Density Sensitivity of Empirical
  Functionals. \emph{J. Phys. Chem. Lett.} \textbf{2021}, \emph{12},
  800--807\relax
\mciteBstWouldAddEndPuncttrue
\mciteSetBstMidEndSepPunct{\mcitedefaultmidpunct}
{\mcitedefaultendpunct}{\mcitedefaultseppunct}\relax
\EndOfBibitem
\bibitem[Kim \latin{et~al.}(2013)Kim, Sim, and Burke]{kim2013}
Kim,~M.-C.; Sim,~E.; Burke,~K. Understanding and Reducing Errors in Density
  Functional Calculations. \emph{Phys. Rev. Lett.} \textbf{2013}, \emph{111},
  073003\relax
\mciteBstWouldAddEndPuncttrue
\mciteSetBstMidEndSepPunct{\mcitedefaultmidpunct}
{\mcitedefaultendpunct}{\mcitedefaultseppunct}\relax
\EndOfBibitem
\bibitem[Martín-Fernández and Harvey(2021)Martín-Fernández, and
  Harvey]{martin2021}
Martín-Fernández,~C.; Harvey,~J.~N. On the Use of Normalized Metrics for
  Density Sensitivity Analysis in DFT. \emph{J. Phys. Chem. A} \textbf{2021},
  \emph{125}, 4639--4652\relax
\mciteBstWouldAddEndPuncttrue
\mciteSetBstMidEndSepPunct{\mcitedefaultmidpunct}
{\mcitedefaultendpunct}{\mcitedefaultseppunct}\relax
\EndOfBibitem
\bibitem[Kim \latin{et~al.}(2019)Kim, Song, Sim, and Burke]{kim2019}
Kim,~Y.; Song,~S.; Sim,~E.; Burke,~K. Halogen and Chalcogen Binding Dominated
  by Density-Driven Errors. \emph{J. Phys. Chem. Lett.} \textbf{2019},
  \emph{10}, 295--301\relax
\mciteBstWouldAddEndPuncttrue
\mciteSetBstMidEndSepPunct{\mcitedefaultmidpunct}
{\mcitedefaultendpunct}{\mcitedefaultseppunct}\relax
\EndOfBibitem
\bibitem[Kim \latin{et~al.}(2014)Kim, Sim, and Burke]{kim2014}
Kim,~M.-C.; Sim,~E.; Burke,~K. Ions in solution: Density corrected density
  functional theory (DC-DFT). \emph{J. Chem. Phys.} \textbf{2014}, \emph{140},
  18A528\relax
\mciteBstWouldAddEndPuncttrue
\mciteSetBstMidEndSepPunct{\mcitedefaultmidpunct}
{\mcitedefaultendpunct}{\mcitedefaultseppunct}\relax
\EndOfBibitem
\bibitem[Wasserman \latin{et~al.}(2017)Wasserman, Nafziger, Jiang, Kim, Sim,
  and Burke]{wasserman2017}
Wasserman,~A.; Nafziger,~J.; Jiang,~K.; Kim,~M.-C.; Sim,~E.; Burke,~K. The
  Importance of Being Inconsistent. \emph{Annu. Rev. Phys. Chem.}
  \textbf{2017}, \emph{68}, 555--581\relax
\mciteBstWouldAddEndPuncttrue
\mciteSetBstMidEndSepPunct{\mcitedefaultmidpunct}
{\mcitedefaultendpunct}{\mcitedefaultseppunct}\relax
\EndOfBibitem
\bibitem[Kim \latin{et~al.}(2011)Kim, Sim, and Burke]{kim2011}
Kim,~M.-C.; Sim,~E.; Burke,~K. Communication: Avoiding unbound anions in
  density functional calculations. \emph{J. Chem. Phys.} \textbf{2011},
  \emph{134}, 171103\relax
\mciteBstWouldAddEndPuncttrue
\mciteSetBstMidEndSepPunct{\mcitedefaultmidpunct}
{\mcitedefaultendpunct}{\mcitedefaultseppunct}\relax
\EndOfBibitem
\bibitem[Kim \latin{et~al.}(2015)Kim, Park, Son, Sim, and Burke]{kim2015}
Kim,~M.-C.; Park,~H.; Son,~S.; Sim,~E.; Burke,~K. Improved DFT Potential Energy
  Surfaces via Improved Densities. \emph{J. Phys. Chem. Lett.} \textbf{2015},
  \emph{6}, 3802--3807\relax
\mciteBstWouldAddEndPuncttrue
\mciteSetBstMidEndSepPunct{\mcitedefaultmidpunct}
{\mcitedefaultendpunct}{\mcitedefaultseppunct}\relax
\EndOfBibitem
\bibitem[Song \latin{et~al.}(2018)Song, Kim, Sim, Benali, Heinonen, and
  Burke]{song2018}
Song,~S.; Kim,~M.-C.; Sim,~E.; Benali,~A.; Heinonen,~O.; Burke,~K. Benchmarks
  and Reliable DFT Results for Spin Gaps of Small Ligand Fe(II) Complexes.
  \emph{J. Chem. Theory Comput.} \textbf{2018}, \emph{14}, 2304--2311\relax
\mciteBstWouldAddEndPuncttrue
\mciteSetBstMidEndSepPunct{\mcitedefaultmidpunct}
{\mcitedefaultendpunct}{\mcitedefaultseppunct}\relax
\EndOfBibitem
\bibitem[Lee \latin{et~al.}(2010)Lee, Furche, and Burke]{lee2010b}
Lee,~D.; Furche,~F.; Burke,~K. Accuracy of Electron Affinities of Atoms in
  Approximate Density Functional Theory. \emph{J. Phys. Chem. Lett.}
  \textbf{2010}, \emph{1}, 2124--2129\relax
\mciteBstWouldAddEndPuncttrue
\mciteSetBstMidEndSepPunct{\mcitedefaultmidpunct}
{\mcitedefaultendpunct}{\mcitedefaultseppunct}\relax
\EndOfBibitem
\bibitem[Lee and Burke(2010)Lee, and Burke]{lee2010}
Lee,~D.; Burke,~K. Finding electron affinities with approximate density
  functionals. \emph{Mol. Phys.} \textbf{2010}, \emph{108}, 2687--2701\relax
\mciteBstWouldAddEndPuncttrue
\mciteSetBstMidEndSepPunct{\mcitedefaultmidpunct}
{\mcitedefaultendpunct}{\mcitedefaultseppunct}\relax
\EndOfBibitem
\bibitem[Lambros \latin{et~al.}(2021)Lambros, Dasgupta, Palos, Swee, Hu, and
  Paesani]{lambros2021}
Lambros,~E.; Dasgupta,~S.; Palos,~E.; Swee,~S.; Hu,~J.; Paesani,~F. General
  Many-Body Framework for Data-Driven Potentials with Arbitrary Quantum
  Mechanical Accuracy: Water as a Case Study. \emph{J. Chem. Theory Comput.}
  \textbf{2021}, \emph{17}, 5635--5650\relax
\mciteBstWouldAddEndPuncttrue
\mciteSetBstMidEndSepPunct{\mcitedefaultmidpunct}
{\mcitedefaultendpunct}{\mcitedefaultseppunct}\relax
\EndOfBibitem
\bibitem[Dasgupta \latin{et~al.}(2021)Dasgupta, Lambros, Perdew, and
  Paesani]{dasgupta2021}
Dasgupta,~S.; Lambros,~E.; Perdew,~J.~P.; Paesani,~F. Elevating density
  functional theory to chemical accuracy for water simulations through a
  density-corrected many-body formalism. \emph{Nat. Commun.} \textbf{2021},
  \emph{12}, 6359\relax
\mciteBstWouldAddEndPuncttrue
\mciteSetBstMidEndSepPunct{\mcitedefaultmidpunct}
{\mcitedefaultendpunct}{\mcitedefaultseppunct}\relax
\EndOfBibitem
\bibitem[Umrigar and Gonze(1994)Umrigar, and Gonze]{umrigar1994}
Umrigar,~C.~J.; Gonze,~X. Accurate exchange-correlation potentials and
  total-energy components for the helium isoelectronic series. \emph{Phys. Rev.
  A} \textbf{1994}, \emph{50}, 3827--3837\relax
\mciteBstWouldAddEndPuncttrue
\mciteSetBstMidEndSepPunct{\mcitedefaultmidpunct}
{\mcitedefaultendpunct}{\mcitedefaultseppunct}\relax
\EndOfBibitem
\bibitem[Cruz \latin{et~al.}(1998)Cruz, Lam, and Burke]{cruz1998}
Cruz,~F.~G.; Lam,~K.-C.; Burke,~K. Exchange-Correlation Energy Density from
  Virial Theorem. \emph{J. Phys. Chem. A} \textbf{1998}, \emph{102},
  4911--4917\relax
\mciteBstWouldAddEndPuncttrue
\mciteSetBstMidEndSepPunct{\mcitedefaultmidpunct}
{\mcitedefaultendpunct}{\mcitedefaultseppunct}\relax
\EndOfBibitem
\bibitem[Kümmel and Kronik(2008)Kümmel, and Kronik]{kuemmel2008}
Kümmel,~S.; Kronik,~L. Orbital-dependent density functionals: Theory and
  applications. \emph{Rev. Mod. Phys.} \textbf{2008}, \emph{80}, 3--60\relax
\mciteBstWouldAddEndPuncttrue
\mciteSetBstMidEndSepPunct{\mcitedefaultmidpunct}
{\mcitedefaultendpunct}{\mcitedefaultseppunct}\relax
\EndOfBibitem
\bibitem[Burke \latin{et~al.}(1998)Burke, Cruz, and Lam]{burke1998}
Burke,~K.; Cruz,~F.~G.; Lam,~K.-C. Unambiguous exchange-correlation energy
  density. \emph{J. Chem. Phys.} \textbf{1998}, \emph{109}, 8161--8167\relax
\mciteBstWouldAddEndPuncttrue
\mciteSetBstMidEndSepPunct{\mcitedefaultmidpunct}
{\mcitedefaultendpunct}{\mcitedefaultseppunct}\relax
\EndOfBibitem
\bibitem[Kaplan \latin{et~al.}(2023)Kaplan, Shahi, Bhetwal, Sah, and
  Perdew]{kaplan2023}
Kaplan,~A.~D.; Shahi,~C.; Bhetwal,~P.; Sah,~R.~K.; Perdew,~J.~P. Understanding
  Density-Driven Errors for Reaction Barrier Heights. \emph{J. Chem. Theory
  Comput.} \textbf{2023}, \emph{19}, 532--543\relax
\mciteBstWouldAddEndPuncttrue
\mciteSetBstMidEndSepPunct{\mcitedefaultmidpunct}
{\mcitedefaultendpunct}{\mcitedefaultseppunct}\relax
\EndOfBibitem
\bibitem[Mezei \latin{et~al.}(2017)Mezei, Csonka, and Kállay]{mezei2017}
Mezei,~P.~D.; Csonka,~G.~I.; Kállay,~M. Electron Density Errors and
  Density-Driven Exchange-Correlation Energy Errors in Approximate Density
  Functional Calculations. \emph{J. Chem. Theory Comput.} \textbf{2017},
  \emph{13}, 4753--4764\relax
\mciteBstWouldAddEndPuncttrue
\mciteSetBstMidEndSepPunct{\mcitedefaultmidpunct}
{\mcitedefaultendpunct}{\mcitedefaultseppunct}\relax
\EndOfBibitem
\bibitem[Medvedev \latin{et~al.}(2017)Medvedev, Bushmarinov, Sun, Perdew, and
  Lyssenko]{medvedev2017}
Medvedev,~M.~G.; Bushmarinov,~I.~S.; Sun,~J.; Perdew,~J.~P.; Lyssenko,~K.~A.
  Density functional theory is straying from the path toward the exact
  functional. \emph{Science} \textbf{2017}, \emph{355}, 49--52\relax
\mciteBstWouldAddEndPuncttrue
\mciteSetBstMidEndSepPunct{\mcitedefaultmidpunct}
{\mcitedefaultendpunct}{\mcitedefaultseppunct}\relax
\EndOfBibitem
\bibitem[Brorsen \latin{et~al.}(2017)Brorsen, Yang, Pak, and
  Hammes-Schiffer]{brorsen2017}
Brorsen,~K.~R.; Yang,~Y.; Pak,~M.~V.; Hammes-Schiffer,~S. Is the Accuracy of
  Density Functional Theory for Atomization Energies and Densities in Bonding
  Regions Correlated? \emph{J. Phys. Chem. Lett.} \textbf{2017}, \emph{8},
  2076--2081\relax
\mciteBstWouldAddEndPuncttrue
\mciteSetBstMidEndSepPunct{\mcitedefaultmidpunct}
{\mcitedefaultendpunct}{\mcitedefaultseppunct}\relax
\EndOfBibitem
\bibitem[Mostafanejad \latin{et~al.}(2019)Mostafanejad, Haney, and
  DePrince]{mostafanejad2019}
Mostafanejad,~M.; Haney,~J.; DePrince,~A.~E. Kinetic-energy-based error
  quantification in Kohn–Sham density functional theory. \emph{Phys. Chem.
  Chem. Phys.} \textbf{2019}, \emph{21}, 26492--26501\relax
\mciteBstWouldAddEndPuncttrue
\mciteSetBstMidEndSepPunct{\mcitedefaultmidpunct}
{\mcitedefaultendpunct}{\mcitedefaultseppunct}\relax
\EndOfBibitem
\bibitem[Medvedev \latin{et~al.}(2017)Medvedev, Bushmarinov, Sun, Perdew, and
  Lyssenko]{medvedev2017b}
Medvedev,~M.~G.; Bushmarinov,~I.~S.; Sun,~J.; Perdew,~J.~P.; Lyssenko,~K.~A.
  Response to Comment on ``Density functional theory is straying from the path
  toward the exact functional’’. \emph{Science} \textbf{2017}, \emph{356},
  496--496\relax
\mciteBstWouldAddEndPuncttrue
\mciteSetBstMidEndSepPunct{\mcitedefaultmidpunct}
{\mcitedefaultendpunct}{\mcitedefaultseppunct}\relax
\EndOfBibitem
\bibitem[Hammes-Schiffer(2017)]{hammes2017}
Hammes-Schiffer,~S. A conundrum for density functional theory. \emph{Science}
  \textbf{2017}, \emph{355}, 28--29\relax
\mciteBstWouldAddEndPuncttrue
\mciteSetBstMidEndSepPunct{\mcitedefaultmidpunct}
{\mcitedefaultendpunct}{\mcitedefaultseppunct}\relax
\EndOfBibitem
\bibitem[Kepp(2017)]{kepp2017}
Kepp,~K.~P. Comment on ``Density functional theory is straying from the path
  toward the exact functional’’. \emph{Science} \textbf{2017}, \emph{356},
  496--496\relax
\mciteBstWouldAddEndPuncttrue
\mciteSetBstMidEndSepPunct{\mcitedefaultmidpunct}
{\mcitedefaultendpunct}{\mcitedefaultseppunct}\relax
\EndOfBibitem
\bibitem[Gould(2017)]{gould2017}
Gould,~T. What Makes a Density Functional Approximation Good? Insights from the
  Left Fukui Function. \emph{J. Chem. Theory Comput.} \textbf{2017}, \emph{13},
  2373--2377\relax
\mciteBstWouldAddEndPuncttrue
\mciteSetBstMidEndSepPunct{\mcitedefaultmidpunct}
{\mcitedefaultendpunct}{\mcitedefaultseppunct}\relax
\EndOfBibitem
\bibitem[Mayer \latin{et~al.}(2017)Mayer, P\'{a}pai, Bak\'{o}, and
  Nagy]{mayer2017}
Mayer,~I.; P\'{a}pai,~I.; Bak\'{o},~I.; Nagy,~A. Conceptual Problem with
  Calculating Electron Densities in Finite Basis Density Functional Theory.
  \emph{J. Chem. Theory Comput.} \textbf{2017}, \emph{13}, 3961--3963\relax
\mciteBstWouldAddEndPuncttrue
\mciteSetBstMidEndSepPunct{\mcitedefaultmidpunct}
{\mcitedefaultendpunct}{\mcitedefaultseppunct}\relax
\EndOfBibitem
\bibitem[Song \latin{et~al.}(2022)Song, Vuckovic, Sim, and Burke]{song2022}
Song,~S.; Vuckovic,~S.; Sim,~E.; Burke,~K. Density-Corrected DFT Explained:
  Questions and Answers. \emph{J. Chem. Theory Comput.} \textbf{2022},
  \emph{18}, 817--827\relax
\mciteBstWouldAddEndPuncttrue
\mciteSetBstMidEndSepPunct{\mcitedefaultmidpunct}
{\mcitedefaultendpunct}{\mcitedefaultseppunct}\relax
\EndOfBibitem
\bibitem[Kepp(2018)]{kepp2018}
Kepp,~K.~P. Energy vs. density on paths toward more exact density functionals.
  \emph{Phys. Chem. Chem. Phys.} \textbf{2018}, \emph{20}, 7538--7548\relax
\mciteBstWouldAddEndPuncttrue
\mciteSetBstMidEndSepPunct{\mcitedefaultmidpunct}
{\mcitedefaultendpunct}{\mcitedefaultseppunct}\relax
\EndOfBibitem
\bibitem[Crisostomo \latin{et~al.}(2022)Crisostomo, Pederson, Kozlowski,
  Kalita, Cancio, Datchev, Wasserman, Song, and Burke]{crisostomo2022b}
Crisostomo,~S.; Pederson,~R.; Kozlowski,~J.; Kalita,~B.; Cancio,~A.~C.;
  Datchev,~K.; Wasserman,~A.; Song,~S.; Burke,~K. Seven Useful Questions in
  Density Functional Theory. \emph{arXiv preprint arXiv:2207.05794}
  \textbf{2022}, \relax
\mciteBstWouldAddEndPunctfalse
\mciteSetBstMidEndSepPunct{\mcitedefaultmidpunct}
{}{\mcitedefaultseppunct}\relax
\EndOfBibitem
\bibitem[Mart\'{i}n~Pend\'{a}s and Francisco(2022)Mart\'{i}n~Pend\'{a}s, and
  Francisco]{pendas2022}
Mart\'{i}n~Pend\'{a}s,~A.; Francisco,~E. The role of references and the elusive
  nature of the chemical bond. \emph{Nat. Commun.} \textbf{2022}, \emph{13},
  3327\relax
\mciteBstWouldAddEndPuncttrue
\mciteSetBstMidEndSepPunct{\mcitedefaultmidpunct}
{\mcitedefaultendpunct}{\mcitedefaultseppunct}\relax
\EndOfBibitem
\bibitem[Perdew \latin{et~al.}(1996)Perdew, Burke, and Ernzerhof]{perdew1996}
Perdew,~J.~P.; Burke,~K.; Ernzerhof,~M. Generalized Gradient Approximation Made
  Simple. \emph{Phys. Rev. Lett.} \textbf{1996}, \emph{77}, 3865--3868\relax
\mciteBstWouldAddEndPuncttrue
\mciteSetBstMidEndSepPunct{\mcitedefaultmidpunct}
{\mcitedefaultendpunct}{\mcitedefaultseppunct}\relax
\EndOfBibitem
\bibitem[Perdew \latin{et~al.}(1997)Perdew, Burke, and Ernzerhof]{perdew1997}
Perdew,~J.~P.; Burke,~K.; Ernzerhof,~M. Generalized Gradient Approximation Made
  Simple [Phys. Rev. Lett. 77, 3865 (1996)]. \emph{Phys. Rev. Lett.}
  \textbf{1997}, \emph{78}, 1396--1396\relax
\mciteBstWouldAddEndPuncttrue
\mciteSetBstMidEndSepPunct{\mcitedefaultmidpunct}
{\mcitedefaultendpunct}{\mcitedefaultseppunct}\relax
\EndOfBibitem
\bibitem[Adamo and Barone(1999)Adamo, and Barone]{adamo1999}
Adamo,~C.; Barone,~V. Toward reliable density functional methods without
  adjustable parameters: The PBE0 model. \emph{J. Chem. Phys.} \textbf{1999},
  \emph{110}, 6158--6170\relax
\mciteBstWouldAddEndPuncttrue
\mciteSetBstMidEndSepPunct{\mcitedefaultmidpunct}
{\mcitedefaultendpunct}{\mcitedefaultseppunct}\relax
\EndOfBibitem
\bibitem[Ernzerhof and Scuseria(1999)Ernzerhof, and Scuseria]{ernzerhof1999}
Ernzerhof,~M.; Scuseria,~G.~E. Assessment of the Perdew–Burke–Ernzerhof
  exchange-correlation functional. \emph{J. Chem. Phys.} \textbf{1999},
  \emph{110}, 5029--5036\relax
\mciteBstWouldAddEndPuncttrue
\mciteSetBstMidEndSepPunct{\mcitedefaultmidpunct}
{\mcitedefaultendpunct}{\mcitedefaultseppunct}\relax
\EndOfBibitem
\bibitem[Rodríguez \latin{et~al.}(2009)Rodríguez, Ayers, Götz, and
  Castillo-Alvarado]{rodriguez2009}
Rodríguez,~J.~I.; Ayers,~P.~W.; Götz,~A.~W.; Castillo-Alvarado,~F.~L. Virial
  theorem in the Kohn–Sham density-functional theory formalism: Accurate
  calculation of the atomic quantum theory of atoms in molecules energies.
  \emph{J. Chem. Phys.} \textbf{2009}, \emph{131}, 021101\relax
\mciteBstWouldAddEndPuncttrue
\mciteSetBstMidEndSepPunct{\mcitedefaultmidpunct}
{\mcitedefaultendpunct}{\mcitedefaultseppunct}\relax
\EndOfBibitem
\bibitem[Levy and Perdew(1985)Levy, and Perdew]{levy1985}
Levy,~M.; Perdew,~J.~P. Hellmann-Feynman, virial, and scaling requisites for
  the exact universal density functionals. Shape of the correlation potential
  and diamagnetic susceptibility for atoms. \emph{Phys. Rev. A} \textbf{1985},
  \emph{32}, 2010--2021\relax
\mciteBstWouldAddEndPuncttrue
\mciteSetBstMidEndSepPunct{\mcitedefaultmidpunct}
{\mcitedefaultendpunct}{\mcitedefaultseppunct}\relax
\EndOfBibitem
\bibitem[Gritsenko and Baerends(1996)Gritsenko, and Baerends]{gritsenko1996}
Gritsenko,~O.~V.; Baerends,~E.~J. Effect of molecular dissociation on the
  exchange-correlation Kohn-Sham potential. \emph{Phys. Rev. A} \textbf{1996},
  \emph{54}, 1957--1972\relax
\mciteBstWouldAddEndPuncttrue
\mciteSetBstMidEndSepPunct{\mcitedefaultmidpunct}
{\mcitedefaultendpunct}{\mcitedefaultseppunct}\relax
\EndOfBibitem
\bibitem[Görling and Ernzerhof(1995)Görling, and Ernzerhof]{goerling1995}
Görling,~A.; Ernzerhof,~M. Energy differences between Kohn-Sham and
  Hartree-Fock wave functions yielding the same electron density. \emph{Phys.
  Rev. A} \textbf{1995}, \emph{51}, 4501--4513\relax
\mciteBstWouldAddEndPuncttrue
\mciteSetBstMidEndSepPunct{\mcitedefaultmidpunct}
{\mcitedefaultendpunct}{\mcitedefaultseppunct}\relax
\EndOfBibitem
\bibitem[Crisostomo \latin{et~al.}(2022)Crisostomo, Levy, and
  Burke]{crisostomo2022}
Crisostomo,~S.; Levy,~M.; Burke,~K. Can the Hartree–Fock kinetic energy
  exceed the exact kinetic energy? \emph{J. Chem. Phys.} \textbf{2022},
  \emph{157}, 154106\relax
\mciteBstWouldAddEndPuncttrue
\mciteSetBstMidEndSepPunct{\mcitedefaultmidpunct}
{\mcitedefaultendpunct}{\mcitedefaultseppunct}\relax
\EndOfBibitem
\bibitem[Jure\^{c}ka \latin{et~al.}(2006)Jure\^{c}ka, \^{S}poner,
  \^{C}ern\'{y}, and Hobza]{jurecka2006}
Jure\^{c}ka,~P.; \^{S}poner,~J.; \^{C}ern\'{y},~J.; Hobza,~P. Benchmark
  database of accurate (MP2 and CCSD(T) complete basis set limit) interaction
  energies of small model complexes, DNA base pairs, and amino acid pairs.
  \emph{Phys. Chem. Chem. Phys.} \textbf{2006}, \emph{8}, 1985--1993\relax
\mciteBstWouldAddEndPuncttrue
\mciteSetBstMidEndSepPunct{\mcitedefaultmidpunct}
{\mcitedefaultendpunct}{\mcitedefaultseppunct}\relax
\EndOfBibitem
\bibitem[Marshall \latin{et~al.}(2011)Marshall, Burns, and
  Sherrill]{marshall2011}
Marshall,~M.~S.; Burns,~L.~A.; Sherrill,~C.~D. Basis set convergence of the
  coupled-cluster correction, $\delta$MP2CCSD(T): Best practices for
  benchmarking non-covalent interactions and the attendant revision of the S22,
  NBC10, HBC6, and HSG databases. \emph{J. Chem. Phys.} \textbf{2011},
  \emph{135}, 194102\relax
\mciteBstWouldAddEndPuncttrue
\mciteSetBstMidEndSepPunct{\mcitedefaultmidpunct}
{\mcitedefaultendpunct}{\mcitedefaultseppunct}\relax
\EndOfBibitem
\bibitem[Bauz\'{a} \latin{et~al.}(2013)Bauz\'{a}, Alkorta, Frontera, and
  Elguero]{bauza2013}
Bauz\'{a},~A.; Alkorta,~I.; Frontera,~A.; Elguero,~J. On the Reliability of
  Pure and Hybrid DFT Methods for the Evaluation of Halogen, Chalcogen, and
  Pnicogen Bonds Involving Anionic and Neutral Electron Donors. \emph{J. Chem.
  Theory Comput.} \textbf{2013}, \emph{9}, 5201--5210\relax
\mciteBstWouldAddEndPuncttrue
\mciteSetBstMidEndSepPunct{\mcitedefaultmidpunct}
{\mcitedefaultendpunct}{\mcitedefaultseppunct}\relax
\EndOfBibitem
\bibitem[Otero-de-la Roza \latin{et~al.}(2014)Otero-de-la Roza, Johnson, and
  DiLabio]{roza2014}
Otero-de-la Roza,~A.; Johnson,~E.~R.; DiLabio,~G.~A. Halogen Bonding from
  Dispersion-Corrected Density-Functional Theory: The Role of Delocalization
  Error. \emph{J. Chem. Theory Comput.} \textbf{2014}, \emph{10},
  5436--5447\relax
\mciteBstWouldAddEndPuncttrue
\mciteSetBstMidEndSepPunct{\mcitedefaultmidpunct}
{\mcitedefaultendpunct}{\mcitedefaultseppunct}\relax
\EndOfBibitem
\bibitem[Bloch(1929)]{bloch1929}
Bloch,~F. Bemerkung zur Elektronentheorie des Ferromagnetismus und der
  elektrischen Leitfähigkeit. \emph{Z. Physik} \textbf{1929}, \emph{57},
  545--555\relax
\mciteBstWouldAddEndPuncttrue
\mciteSetBstMidEndSepPunct{\mcitedefaultmidpunct}
{\mcitedefaultendpunct}{\mcitedefaultseppunct}\relax
\EndOfBibitem
\bibitem[Dirac(1930)]{dirac1930}
Dirac,~P. A.~M. Note on Exchange Phenomena in the Thomas Atom. \emph{Math.
  Proc. Cambridge Philos.} \textbf{1930}, \emph{26}, 376--385\relax
\mciteBstWouldAddEndPuncttrue
\mciteSetBstMidEndSepPunct{\mcitedefaultmidpunct}
{\mcitedefaultendpunct}{\mcitedefaultseppunct}\relax
\EndOfBibitem
\bibitem[Vosko \latin{et~al.}(1980)Vosko, Wilk, and Nusair]{vosko1980}
Vosko,~S.~H.; Wilk,~L.; Nusair,~M. Accurate spin-dependent electron liquid
  correlation energies for local spin density calculations: a critical
  analysis. \emph{Can. J. Phys.} \textbf{1980}, \emph{58}, 1200--1211\relax
\mciteBstWouldAddEndPuncttrue
\mciteSetBstMidEndSepPunct{\mcitedefaultmidpunct}
{\mcitedefaultendpunct}{\mcitedefaultseppunct}\relax
\EndOfBibitem
\bibitem[Brandenburg \latin{et~al.}(2016)Brandenburg, Bates, Sun, and
  Perdew]{brandenburg2016}
Brandenburg,~J.~G.; Bates,~J.~E.; Sun,~J.; Perdew,~J.~P. Benchmark tests of a
  strongly constrained semilocal functional with a long-range dispersion
  correction. \emph{Phys. Rev. B} \textbf{2016}, \emph{94}, 115144\relax
\mciteBstWouldAddEndPuncttrue
\mciteSetBstMidEndSepPunct{\mcitedefaultmidpunct}
{\mcitedefaultendpunct}{\mcitedefaultseppunct}\relax
\EndOfBibitem
\bibitem[Furness \latin{et~al.}(2022)Furness, Kaplan, Ning, Perdew, and
  Sun]{furness2022}
Furness,~J.~W.; Kaplan,~A.~D.; Ning,~J.; Perdew,~J.~P.; Sun,~J. Construction of
  meta-GGA functionals through restoration of exact constraint adherence to
  regularized SCAN functionals. \emph{J. Chem. Phys.} \textbf{2022},
  \emph{156}, 034109\relax
\mciteBstWouldAddEndPuncttrue
\mciteSetBstMidEndSepPunct{\mcitedefaultmidpunct}
{\mcitedefaultendpunct}{\mcitedefaultseppunct}\relax
\EndOfBibitem
\bibitem[Zhao and Truhlar(2006)Zhao, and Truhlar]{zhao2006}
Zhao,~Y.; Truhlar,~D.~G. A new local density functional for main-group
  thermochemistry, transition metal bonding, thermochemical kinetics, and
  noncovalent interactions. \emph{J. Chem. Phys.} \textbf{2006}, \emph{125},
  194101\relax
\mciteBstWouldAddEndPuncttrue
\mciteSetBstMidEndSepPunct{\mcitedefaultmidpunct}
{\mcitedefaultendpunct}{\mcitedefaultseppunct}\relax
\EndOfBibitem
\bibitem[Zhao and Truhlar(2008)Zhao, and Truhlar]{zhao2008}
Zhao,~Y.; Truhlar,~D.~G. The M06 suite of density functionals for main group
  thermochemistry, thermochemical kinetics, noncovalent interactions, excited
  states, and transition elements: two new functionals and systematic testing
  of four M06-class functionals and 12 other functionals. \emph{Theor. Chem.
  Acc.} \textbf{2008}, \emph{120}, 215--241\relax
\mciteBstWouldAddEndPuncttrue
\mciteSetBstMidEndSepPunct{\mcitedefaultmidpunct}
{\mcitedefaultendpunct}{\mcitedefaultseppunct}\relax
\EndOfBibitem
\bibitem[Jr.(1989)]{dunning1989}
Jr.,~T. H.~D. Gaussian basis sets for use in correlated molecular calculations.
  I. The atoms boron through neon and hydrogen. \emph{J. Chem. Phys.}
  \textbf{1989}, \emph{90}, 1007--1023\relax
\mciteBstWouldAddEndPuncttrue
\mciteSetBstMidEndSepPunct{\mcitedefaultmidpunct}
{\mcitedefaultendpunct}{\mcitedefaultseppunct}\relax
\EndOfBibitem
\bibitem[Woon and Jr.(1994)Woon, and Jr.]{woon1994}
Woon,~D.~E.; Jr.,~T. H.~D. Gaussian basis sets for use in correlated molecular
  calculations. IV. Calculation of static electrical response properties.
  \emph{J. Chem. Phys.} \textbf{1994}, \emph{100}, 2975--2988\relax
\mciteBstWouldAddEndPuncttrue
\mciteSetBstMidEndSepPunct{\mcitedefaultmidpunct}
{\mcitedefaultendpunct}{\mcitedefaultseppunct}\relax
\EndOfBibitem
\bibitem[Prascher \latin{et~al.}(2011)Prascher, Woon, Peterson, Dunning, and
  Wilson]{prascher2011}
Prascher,~B.~P.; Woon,~D.~E.; Peterson,~K.~A.; Dunning,~T.~H.; Wilson,~A.~K.
  Gaussian basis sets for use in correlated molecular calculations. VII.
  Valence, core-valence, and scalar relativistic basis sets for Li, Be, Na, and
  Mg. \emph{Theor. Chem. Acc.} \textbf{2011}, \emph{128}, 69--82\relax
\mciteBstWouldAddEndPuncttrue
\mciteSetBstMidEndSepPunct{\mcitedefaultmidpunct}
{\mcitedefaultendpunct}{\mcitedefaultseppunct}\relax
\EndOfBibitem
\bibitem[Woon and Jr.(1993)Woon, and Jr.]{woon1993}
Woon,~D.~E.; Jr.,~T. H.~D. Gaussian basis sets for use in correlated molecular
  calculations. III. The atoms aluminum through argon. \emph{J. Chem. Phys.}
  \textbf{1993}, \emph{98}, 1358--1371\relax
\mciteBstWouldAddEndPuncttrue
\mciteSetBstMidEndSepPunct{\mcitedefaultmidpunct}
{\mcitedefaultendpunct}{\mcitedefaultseppunct}\relax
\EndOfBibitem
\bibitem[Friedrich and H\"{a}nchen(2013)Friedrich, and
  H\"{a}nchen]{friedrich2013}
Friedrich,~J.; H\"{a}nchen,~J. Incremental CCSD(T)(F12*)|MP2: A Black Box
  Method To Obtain Highly Accurate Reaction Energies. \emph{J. Chem. Theory
  Comput.} \textbf{2013}, \emph{9}, 5381--5394\relax
\mciteBstWouldAddEndPuncttrue
\mciteSetBstMidEndSepPunct{\mcitedefaultmidpunct}
{\mcitedefaultendpunct}{\mcitedefaultseppunct}\relax
\EndOfBibitem
\bibitem[Goerigk \latin{et~al.}(2017)Goerigk, Hansen, Bauer, Ehrlich, Najibi,
  and Grimme]{goerigk2017}
Goerigk,~L.; Hansen,~A.; Bauer,~C.; Ehrlich,~S.; Najibi,~A.; Grimme,~S. A look
  at the density functional theory zoo with the advanced GMTKN55 database for
  general main group thermochemistry, kinetics and noncovalent interactions.
  \emph{Phys. Chem. Chem. Phys.} \textbf{2017}, \emph{19}, 32184--32215\relax
\mciteBstWouldAddEndPuncttrue
\mciteSetBstMidEndSepPunct{\mcitedefaultmidpunct}
{\mcitedefaultendpunct}{\mcitedefaultseppunct}\relax
\EndOfBibitem
\bibitem[Curtiss \latin{et~al.}(1991)Curtiss, Raghavachari, Trucks, and
  Pople]{curtiss1991}
Curtiss,~L.~A.; Raghavachari,~K.; Trucks,~G.~W.; Pople,~J.~A. Gaussian‐2
  theory for molecular energies of first‐ and second‐row compounds.
  \emph{J. Chem. Phys.} \textbf{1991}, \emph{94}, 7221--7230\relax
\mciteBstWouldAddEndPuncttrue
\mciteSetBstMidEndSepPunct{\mcitedefaultmidpunct}
{\mcitedefaultendpunct}{\mcitedefaultseppunct}\relax
\EndOfBibitem
\bibitem[Goerigk and Grimme(2010)Goerigk, and Grimme]{goerigk2010}
Goerigk,~L.; Grimme,~S. A General Database for Main Group Thermochemistry,
  Kinetics, and Noncovalent Interactions -- Assessment of Common and
  Reparameterized (meta-)GGA Density Functionals. \emph{J. Chem. Theory
  Comput.} \textbf{2010}, \emph{6}, 107--126\relax
\mciteBstWouldAddEndPuncttrue
\mciteSetBstMidEndSepPunct{\mcitedefaultmidpunct}
{\mcitedefaultendpunct}{\mcitedefaultseppunct}\relax
\EndOfBibitem
\bibitem[Johnson \latin{et~al.}(2008)Johnson, Mori-Sánchez, Cohen, and
  Yang]{johnson2008}
Johnson,~E.~R.; Mori-Sánchez,~P.; Cohen,~A.~J.; Yang,~W. Delocalization errors
  in density functionals and implications for main-group thermochemistry.
  \emph{J. Chem. Phys.} \textbf{2008}, \emph{129}, 204112\relax
\mciteBstWouldAddEndPuncttrue
\mciteSetBstMidEndSepPunct{\mcitedefaultmidpunct}
{\mcitedefaultendpunct}{\mcitedefaultseppunct}\relax
\EndOfBibitem
\bibitem[Kussmann and Ochsenfeld(2013)Kussmann, and Ochsenfeld]{kussmann2013}
Kussmann,~J.; Ochsenfeld,~C. Pre-selective screening for matrix elements in
  linear-scaling exact exchange calculations. \emph{J. Chem. Phys.}
  \textbf{2013}, \emph{138}, 134114\relax
\mciteBstWouldAddEndPuncttrue
\mciteSetBstMidEndSepPunct{\mcitedefaultmidpunct}
{\mcitedefaultendpunct}{\mcitedefaultseppunct}\relax
\EndOfBibitem
\bibitem[Kussmann and Ochsenfeld(2015)Kussmann, and Ochsenfeld]{kussmann2015}
Kussmann,~J.; Ochsenfeld,~C. Preselective Screening for Linear-Scaling Exact
  Exchange-Gradient Calculations for Graphics Processing Units and General
  Strong-Scaling Massively Parallel Calculations. \emph{J. Chem. Theory
  Comput.} \textbf{2015}, \emph{11}, 918--922\relax
\mciteBstWouldAddEndPuncttrue
\mciteSetBstMidEndSepPunct{\mcitedefaultmidpunct}
{\mcitedefaultendpunct}{\mcitedefaultseppunct}\relax
\EndOfBibitem
\bibitem[Kussmann and Ochsenfeld(2017)Kussmann, and Ochsenfeld]{kussmann2017}
Kussmann,~J.; Ochsenfeld,~C. Hybrid CPU/GPU Integral Engine for Strong-Scaling
  Ab Initio Methods. \emph{J. Chem. Theory Comput.} \textbf{2017}, \emph{13},
  3153--3159\relax
\mciteBstWouldAddEndPuncttrue
\mciteSetBstMidEndSepPunct{\mcitedefaultmidpunct}
{\mcitedefaultendpunct}{\mcitedefaultseppunct}\relax
\EndOfBibitem
\bibitem[Laqua \latin{et~al.}(2018)Laqua, Kussmann, and Ochsenfeld]{laqua2018}
Laqua,~H.; Kussmann,~J.; Ochsenfeld,~C. An improved molecular partitioning
  scheme for numerical quadratures in density functional theory. \emph{J. Chem.
  Phys.} \textbf{2018}, \emph{149}, 204111\relax
\mciteBstWouldAddEndPuncttrue
\mciteSetBstMidEndSepPunct{\mcitedefaultmidpunct}
{\mcitedefaultendpunct}{\mcitedefaultseppunct}\relax
\EndOfBibitem
\bibitem[Kussmann \latin{et~al.}(2021)Kussmann, Laqua, and
  Ochsenfeld]{kussmann2021}
Kussmann,~J.; Laqua,~H.; Ochsenfeld,~C. Highly Efficient Resolution-of-Identity
  Density Functional Theory Calculations on Central and Graphics Processing
  Units. \emph{J. Chem. Theory Comput.} \textbf{2021}, \emph{17},
  1512--1521\relax
\mciteBstWouldAddEndPuncttrue
\mciteSetBstMidEndSepPunct{\mcitedefaultmidpunct}
{\mcitedefaultendpunct}{\mcitedefaultseppunct}\relax
\EndOfBibitem
\bibitem[Laqua \latin{et~al.}(2020)Laqua, Thompson, Kussmann, and
  Ochsenfeld]{laqua2020}
Laqua,~H.; Thompson,~T.~H.; Kussmann,~J.; Ochsenfeld,~C. Highly Efficient,
  Linear-Scaling Seminumerical Exact-Exchange Method for Graphic Processing
  Units. \emph{J. Chem. Theory Comput.} \textbf{2020}, \emph{16},
  1456--1468\relax
\mciteBstWouldAddEndPuncttrue
\mciteSetBstMidEndSepPunct{\mcitedefaultmidpunct}
{\mcitedefaultendpunct}{\mcitedefaultseppunct}\relax
\EndOfBibitem
\bibitem[Weigend \latin{et~al.}(2003)Weigend, Furche, and
  Ahlrichs]{weigend2003}
Weigend,~F.; Furche,~F.; Ahlrichs,~R. Gaussian basis sets of quadruple zeta
  valence quality for atoms H–Kr. \emph{J. Chem. Phys.} \textbf{2003},
  \emph{119}, 12753--12762\relax
\mciteBstWouldAddEndPuncttrue
\mciteSetBstMidEndSepPunct{\mcitedefaultmidpunct}
{\mcitedefaultendpunct}{\mcitedefaultseppunct}\relax
\EndOfBibitem
\bibitem[Weigend and Ahlrichs(2005)Weigend, and Ahlrichs]{weigend2005}
Weigend,~F.; Ahlrichs,~R. Balanced basis sets of split valence, triple zeta
  valence and quadruple zeta valence quality for H to Rn: Design and assessment
  of accuracy. \emph{Phys. Chem. Chem. Phys.} \textbf{2005}, \emph{7},
  3297--3305\relax
\mciteBstWouldAddEndPuncttrue
\mciteSetBstMidEndSepPunct{\mcitedefaultmidpunct}
{\mcitedefaultendpunct}{\mcitedefaultseppunct}\relax
\EndOfBibitem
\bibitem[Rappoport and Furche(2010)Rappoport, and Furche]{rappoport2010}
Rappoport,~D.; Furche,~F. Property-optimized Gaussian basis sets for molecular
  response calculations. \emph{J. Chem. Phys.} \textbf{2010}, \emph{133},
  134105\relax
\mciteBstWouldAddEndPuncttrue
\mciteSetBstMidEndSepPunct{\mcitedefaultmidpunct}
{\mcitedefaultendpunct}{\mcitedefaultseppunct}\relax
\EndOfBibitem
\bibitem[Weigend(2006)]{weigend2006}
Weigend,~F. Accurate Coulomb-fitting basis sets for H to Rn. \emph{Phys. Chem.
  Chem. Phys.} \textbf{2006}, \emph{8}, 1057--1065\relax
\mciteBstWouldAddEndPuncttrue
\mciteSetBstMidEndSepPunct{\mcitedefaultmidpunct}
{\mcitedefaultendpunct}{\mcitedefaultseppunct}\relax
\EndOfBibitem
\bibitem[Kendall \latin{et~al.}(1992)Kendall, Jr., and Harrison]{kendall1992}
Kendall,~R.~A.; Jr.,~T. H.~D.; Harrison,~R.~J. Electron affinities of the
  first‐row atoms revisited. Systematic basis sets and wave functions.
  \emph{J. Chem. Phys.} \textbf{1992}, \emph{96}, 6796--6806\relax
\mciteBstWouldAddEndPuncttrue
\mciteSetBstMidEndSepPunct{\mcitedefaultmidpunct}
{\mcitedefaultendpunct}{\mcitedefaultseppunct}\relax
\EndOfBibitem
\bibitem[Weigend(2002)]{weigend2002}
Weigend,~F. A fully direct RI-HF algorithm: Implementation, optimised auxiliary
  basis sets, demonstration of accuracy and efficiency. \emph{Phys. Chem. Chem.
  Phys.} \textbf{2002}, \emph{4}, 4285--4291\relax
\mciteBstWouldAddEndPuncttrue
\mciteSetBstMidEndSepPunct{\mcitedefaultmidpunct}
{\mcitedefaultendpunct}{\mcitedefaultseppunct}\relax
\EndOfBibitem
\end{mcitethebibliography}
\end{document}